\documentclass[12pt,a4paper]{article}
\usepackage[applemac]{inputenc}
\usepackage[T1]{fontenc}
\usepackage{amsmath,amsfonts,amssymb}
\usepackage[breaklinks=true]{hyperref}
\usepackage{mathtools}
\mathtoolsset{showonlyrefs}
\usepackage{graphicx,epsfig,float}
\usepackage{color,xcolor}
\usepackage{textcomp}
\usepackage{indentfirst}
\usepackage{multirow}
\usepackage{multicol}
\usepackage{enumerate}
\usepackage[font=small]{caption}
\usepackage{adjustbox}
\usepackage{amsthm}
\usepackage{graphics}
\usepackage{pstricks,pstricks-add,pst-math,pst-xkey}
\usepackage{bm}
\usepackage{nccmath}
\usepackage{url}
\usepackage{bbm}
\usepackage{setspace}
\usepackage[left=1in,right=1in,bottom=1.3in,top=1.3in]{geometry}
\usepackage{parskip}
\usepackage{breakcites}
\usepackage{algorithm}
\usepackage{algpseudocode}
\usepackage{tikz}
\usetikzlibrary{automata, positioning}
\usetikzlibrary{matrix,arrows,decorations.pathmorphing,positioning,graphs,calc}
\usepackage{natbib}
\usepackage{subcaption}
\usepackage{subfiles}
\usepackage[title]{appendix}
\usepackage{titling}
\usepackage{mathtools}
\mathtoolsset{showonlyrefs}
\usepackage{placeins}
\usepackage{soul}

\newcommand{\Lcal}{{\mathcal{L}}}
\newcommand{\MVN}{{\text{MVN}}}
\theoremstyle{definition}

\theoremstyle{plain}

\theoremstyle{definition}

\providecommand{\keywords}[1]{{\small{\it Keywords:} #1}}

\begin{document}
\title{Gaussian mixture copulas for flexible dependence modelling in the body and tails of joint distributions}
\author{L. M. Andr\'e$^{1*}$ and J. A. Tawn$^{2}$\\
\small $^{1}$ Namur Institute for Complex Systems, University of Namur, Rue Graf\'e 2, Namur 5000, Belgium\\
\small $^{2}$ School of Mathematical Sciences, Lancaster University, LA1 4YF, United Kingdom \\
\small $^*$ Correspondence to: \href{mailto:lidia.andre@unamur.be}{lidia.andre@unamur.be} }
\date{\today}

\maketitle
\pagenumbering{arabic}

\begin{abstract}
    Fully describing the entire data set is essential in multivariate risk assessment, since moderate levels of one variable can influence another, potentially leading it to be extreme. Additionally, modelling both non-extreme and extreme events within a single framework avoids the need to select a threshold vector used to determine an extremal region, or the requirement to add flexibility to bridge between separate models for the body and tail regions.  We propose a copula model, based on a mixture of Gaussian distributions, as this model avoids the need to define an extremal region, it is scalable to dimensions beyond the bivariate case, and it can handle both asymptotic dependent and asymptotic independent extremal dependence structures. We apply the proposed model through simulations and to a 5-dimensional seasonal air pollution data set, previously analysed in the multivariate extremes literature. Through pairwise, trivariate and 5-dimensional analyses, we show the flexibility of the Gaussian mixture copula in capturing different joint distributional behaviours and its ability to identify potential graphical structure features, both of which can vary across the body and tail regions.
\end{abstract}


\section{Introduction} \label{section:introduction}

Building infrastructure and developing policies for preventing and protecting society from extreme events is of critical importance. Such impactful events include high temperatures, floods, air pollutants, and market crashes. In each of these events the marginal behaviour is a key feature, but their impact is worsened when the dependence between variables exacerbates the effects. For instance, high temperatures over a small region in one event is much less impactful than if the heatwave is widespread, or the market crash causes much larger losses for investors if stocks in multiple sectors drop simultaneously than when such drops are restricted to one sector. We focus on air pollutant concentrations, where there are multiple different types of air pollutants, and the health implications are much worse if raised levels occur over different combinations of these pollutants. In all these examples, solutions require a multivariate distributional approach that focuses on fitting well both the body and the joint tail regions. 

Traditional extreme value analysis approaches define an extreme region on which inference is based on asymptotically justified models. Such models, however, only use an estimate of the probability of the variables falling in the extreme region, while assigning equal weight on the precise values of data in this region for the inference and extrapolation procedures. Observations outside this extreme region are thus ignored for inference, regardless of their proximity to it. Furthermore, the extreme region is defined by a threshold, or a set of thresholds, depending on the framework. In univariate extremes, the body and upper tail regions of a random variable $X$ are defined by the events $\{X<u\}$ and $\{X\geq u\},$ respectively, for a suitably defined threshold $u$ \citep{Coles2001}. When moving to a $d$-dimensional multivariate setting, with variable $\bm X=(X_1, \ldots ,X_d)\in \mathbb{R}^d$, there is no unique way of defining such an extreme region due to the lack of ordering \citep{Barnett1976}. The most common approaches either assume a multivariate regular varying approach with a single threshold $u$, or a multivariate threshold vector $\bm u = (u_1, \ldots, u_d)\in \mathbb{R}^d.$ Whilst for the former, the extreme region is defined as $\sum_{i=1}^d X_i>u,$ where $\bm X$ has common regularly varying marginal \citep{deHaanFerreira2006}, for the latter, this joint extremal region is defined either by the observations that jointly exceed $\bm u$ \citep{LedfordTawn1996}, or by observations that exceed $\bm u$ in at least one component \citep{HeffernanTawn2004}.  

Even if inference of the distribution of $\bm X$ is only required for the extreme region, selecting such thresholds is often viewed as somewhat arbitrary, with generally an inadequate methodological literature and a very limited set of automated tools to use. This choice can be highly sensitive to which observations are treated as extreme, since those very close to the extreme region boundary are handled entirely differently whether they fall just in or out of this region. Moreover, small changes in the value of thresholds potentially lead to substantially, and practically, different outcomes. Additionally, the selection of thresholds requires a trade-off between bias and variance; choosing low values leads to bias, while setting them too high increases variability unnecessarily. Furthermore, the vast majority of such analyses ignore the substantial uncertainty in the threshold selection on the subsequent inferences, thus tending to give poor coverage probabilities for estimates of extreme events \citep{Murphyetal2024}.

The issues of modelling where thresholds are used to select an extreme region are further complicated when there is an equal interest in the behaviour of $\bm X$ in the non-extreme region. The asymptotically justified models for the extreme region are unlikely to be an appropriate fit for non-extreme observations. Thus, the majority of extreme methods use empirical distributions for this region (e.g., \citealp{ColesTawn1994} and \citealp{HeffernanTawn2004}). However, there are problems with making the resulting estimates of joint distributions and density continuous across the thresholds, and empirical dependence structure models do not provide the transparency, interpretability or the ease with which to incorporate covariates, when compared to parametric dependence models. 

The strategy we adopt avoids the choice of thresholds all together; instead we develop parametric models which allow sufficient flexibility in jointly modelling the body and tail regions of the data. Our focus is specifically on the dependence structure of $\bm X.$ Moreover, the proposed approach is at odds with the traditional extreme value modelling perspective that fitting the entire data set with common statistical distributions might sometimes lead to a poor fit of the model to the data in the extremes, as the fitting criteria put less weight on the tail than in the body of the distribution. Regardless, it has a number of potential advantages. For instance, we do not require defining thresholds, avoiding the associated subjectivity. It also consists of a single model, so there is a smooth transition into the joint tail from the body, and information about the model parameter estimates arise from information in the data across whole the joint distribution. Finally, our model set up enables us to investigate graphical model structures for the data in the body and joint tails of the distribution.

Unlike relying on a singular copula family to exhibit the right features for the body and joint tail region, our approach takes a mixture model formulation. In this way, we avoid making decisions on the form of extremal dependence required to be modelled prior to the analysis. Specifically, and as discussed in Section~\ref{subsection:extremaldeptypes}, there are two types of possible extremal dependence in the bivariate case: asymptotic dependence (AD) and asymptotic independence (AI). All the standard copulas capture only one or the other, and hence do not have the required flexibility as a generic model for extreme regions. Thus, we propose a copula model constructed from a mixture of multivariate Gaussian distributions, overcoming the limitations of the existing approaches. In particular, it accommodates both AI and AD, while avoiding the selection of a threshold vector $\bm u$ and, subsequently, the need for defining an extreme region. In addition, we only need to specify the number of Gaussian mixture components to incorporate in the model, for which we develop diagnostics tools to guide this choice. Therefore, while avoiding the choice of copulas or distributions to take, this copula model is suitable to fit the two regimes of extremal dependence, and is fast to evaluate, scaling relatively well to dimensions $d\geq 2$. 

Considering mixture formulations for merging information from the body and the extreme region of a distribution is not new. In the univariate case, mixture model formulations use asymptotic theory to justify the use of the generalised Pareto (GP) distribution to the excesses of a high threshold $u$ \citep{DavisonSmith1990}, and in some cases a different GP model for the lower tail. \citet{ScarrottMacDonald2012} review univariate models that fit a mixture of the GP distribution and a different parametric model, more suitable to the body of the distribution; we refer to these as extreme value mixture models (EVMMs). Core examples of such models include a dynamically weighted mixture model proposed by \citet{Frigessietal2002}; the models of \citet{Behrensetal2004}, \citet{CarreauBengio2009}, \citet{Tancredietal2006} and \citet{MacDonaldetal2011}, which fit the bulk region in a parametric, semi-parametric or non-parametric way with the threshold $u$ treated as a parameter; the extended GP distribution proposed by \cite{PapastathopoulosTawn2013} and \cite{Naveauetal2016}, whereby the threshold is removed by embedding the GP model into a broader parametric family with the flexibility of the GP tail formulation; and hierarchical models proposed by \citet{CastroCamiloetal2019} and \citet{Yadavetal2021}, which splice Gamma distributions to tail models for extra flexibility. By implicitly or explicitly treating the threshold $u$ as a parameter of the model, the majority of these models aim to account for the uncertainty around this threshold choice in the inference procedure. 

In bivariate problems, mixture model formulations are typically expressed via copulas, i.e., after componentwise transformation of $\bm X$ to uniformly distributed variables $\bm U\in [0,1]^d$ \citep{Joe2014}. In \citet{Aulbachetal2012, Aulbachetal2012b}, one copula is fitted for the bulk region defined by the observations $\{\bm U \not\leq \bm u\},$  and the copula of a bivariate extreme value copula, is used for the extreme region $\{\bm U> \bm u\},$ \citet{LeonelliGamerman2020} propose a mixture of elliptical copulas, which are fit to the full range of the data, and \citet{Andreetal2024} extend the model of \citet{Frigessietal2002} to a bivariate setting, whereby two copulas are fitted to the full support of the data and are  blended through a non-decreasing dynamic weighting function. These copula models do not necessarily cover both types of extremal dependence (i.e., AD and AI) even in the bivariate case. Specifically, the model from \citet{Aulbachetal2012,Aulbachetal2012b} is only suitable for modelling AD data, whereas the models from \citet{LeonelliGamerman2020} and \citet{Andreetal2024} capture both regimes, although the former can only capture AI if all the copulas in the mixture exhibit AI. This restriction of the \citet{LeonelliGamerman2020} approach makes the analysis highly sensitive as it is necessary to determine which extremal dependence regime is appropriate to fit the data prior to fitting their model. Finally, \citet{LeonelliGamerman2020} and \citet{Andreetal2024} require choosing a priori which copula families to include in the mixture model, which then requires multiple model fits using a range of copula combinations. 

This paper is organised as follows: in Section~\ref{section:model} we review key measures for modelling the extremal dependence of a data set, define our proposed model, highlight its extremal dependence properties, and outline the inference and diagnostic procedures. Section~\ref{section:modelfit} presents simulation studies performed to assess the performance of the model. We apply our methodology on the $5$-dimensional seasonal air pollution data set analysed by \citet{HeffernanTawn2004} in Section~\ref{section:casestudy} and conclude in Section~\ref{section:conclusion}.
 
\section{Methodology} \label{section:model}

\subsection{Introduction to extremal dependence modelling}
\label{subsection:extremaldeptypes}
Let $\bm X = (X_1, \ldots, X_d)$ be a $d$-dimensional random vector with $d\geq 2$ and $X_i$ be the $i^{th}$ marginal variables for $i\in D =\{1,\ldots, d\}.$ Sklar's theorem \citep{Sklar1959} states that, if $\bm X$ has joint distribution function $F_{\bm X},$ marginal distribution functions $F_{X_i}$ $(i\in D),$ and is a jointly continuous variable, then there exists a unique copula $C_{\bm X}:[0,1]^d \to [0,1]$ such that, for all $(u_1, \ldots, u_d)\in [0,1]^d,$ 
\begin{equation}\label{eq:sklar}
    C_{\bm X}(u_1, \ldots, u_d) = F_{\bm X}\left(F_{X_1}^{-1}(u_1), \ldots, F_{X_d}^{-1}(u_d)\right).
\end{equation}
The dependence structure of $\bm X$ is then fully captured through the copula $C_{\bm X},$ independently of the margins. Additionally, where it exists, the copula density $c_{\bm X}(u_1, \ldots, u_d)$ can be obtained by taking the $d^{th}$ order mixed derivative of $C_{\bm X}$ with respect to the variables $u_i,$ $i \in D.$ A review of a range of parametric copulas is provided by \citet{Joe1997}, and methods for construction of flexible copula parametric families, which are tailored to capture different dependence structures throughout the support $[0,1]^d$ with $d\geq 2,$ are reviewed by \citet{Andreetal2024}. 

In multivariate extremes, capturing the extremal behaviour of a data set is key to correctly analyse and draw appropriate conclusions about the data set in hand. In particular, we are interested in models that are flexible enough to accommodate the two regimes of extremal dependence: asymptotic dependence (AD), where the variables are likely to occur together at an extreme level, or asymptotic independence (AI), otherwise. The joint tail behaviour of a random vector $\bm X$ can be quantified through the measure $\chi_D$ \citep{Joe1997, Colesetal1999}, which is defined, where it exists, via the limit \linebreak $\chi_D = \lim_{r\to 1}\chi_D(r) \in [0, 1]$ with
\begin{equation}\label{eq:chiddim}
    \chi_D(r)=\Pr[F_{X_i}(X_i) > r: \forall i \in D\setminus \{1\} \mid F_{X_1}(X_1) > r] =\frac{\Pr[F_{X_i}(X_i) > r: \forall i \in D]}{1-r},
\end{equation}
for $r\in (0,1).$ If $\chi_D > 0,$ the variables in $\bm X$ are said to be AD, whilst in the case where $\chi_D =0,$ the variables cannot take all their largest values together. Moreover, larger values of $\chi_D$ indicate stronger AD levels. In the bivariate case, when $\chi_D=0$, it can be said that $X_1$ and $X_2$ are AI; however, care is needed in higher dimensions as a lower dimensional subvector $\bm X_C=\{X_i:i \in C\}$, where $C \subset D,$ could still exhibit AD and have $\chi_C>0$ even though $\chi_D=0$ \citep{Simpsonetal2020}.

A complementary measure to $\chi_D$ was introduced by \citet{LedfordTawn1996} in the bivariate case, and presented in $d$-dimensions by \citet{EastoeTawn2012}. Given a function $\mathcal{L}_D$ that is slowly-varying at infinity, the joint tail of $\bm X$ can be characterised as
\begin{equation}\label{eq:etaddim}
    \Pr[F_{X_i}(X_i) > r: \forall i \in D \setminus \{1\} \mid F_{X_1}(X_1) > r] \sim \mathcal{L}_D((1-r)^{-1})(1-r)^{1\slash \eta_D - 1},
\end{equation}
as $r\to 1,$ and $\eta_D \in (0, 1].$ The extremal dependence structure is quantified through the measure $\eta_D;$ when $\eta_D = 1$ and $\mathcal{L}_D(x)\not\to 0$ as $x\to \infty,$ then the variables in $\bm X$ are AD, and if $\eta_D < 1,$ then they cannot all be extreme together. When $d=2,$ the vector $(X_1,X_2)$ is AI in the latter case. Furthermore, the coefficient $\eta_D$ provides insight about the strength of AI of a given random vector. In particular, if $\eta_D = 1\slash d$ and $\Lcal_D(x) = 1$ $(\Lcal_D(x) \neq 1)$ then independence (near independence) is achieved, whereas when $\eta_D > 1\slash d$ $(\eta_D < 1 \slash d),$ there is evidence of positive (negative) dependence in the extremes. Similar to $\chi_D,$ the coefficient $\eta_D$ is taken as the limit $\eta_D = \lim_{r\to 1} \eta_D(r),$ where it exists, with
\begin{equation}\label{eq:etarddim}
    \eta_D(r)= \frac{\log(1-r)}{\log\left(\Pr[F_{X_i}(X_i) > r: \forall i \in D]\right)},\quad r\in(0,1).
\end{equation}

The two measures $(\chi_D, \eta_D)$ of extremal dependence, however, are only informative when studying the joint tail behaviour, i.e., when all the variables are extreme together. An extension of expression~\eqref{eq:etaddim} was proposed by \citet{WadsworthTawn2013} for when the interest lies instead in regions where variables are not required to be equally extreme over different margins. Specifically, the relative level of extremity across $\bm X$ is represented by $\bm w = (w_1, \ldots, w_d) \in \mathcal{S}_{d-1}:=\{\bm w\in [0,1]^d:\sum_{i=1}^d w_i = 1\}$. In this approach, the joint tail behaviour of $\bm X$ is captured through the function $\lambda_D(\bm w)$ via the assumption that, for any $\bm w\in \mathcal{S}_{d-1}$ with $w_t>0,$ for any selected $t\in D,$ we then have
\begin{equation}\label{eq:lambdaddim}
    \Pr[F_{X_i}(X_i) > 1-(1-r)^{w_i\slash w_t}: \forall i \in D] \sim \Lcal_{\bm w}[(1-r)^{-1\slash w_t}](1-r)^{\lambda_D(\bm w)\slash w_t} 
\end{equation}
as $r\to 1,$ where function $\Lcal_{\bm w}[(1-r)^{-1\slash w_t}]$ is slowly-varying at infinity (implying that $\Lcal_{\bm w}(x)$ is slowly-varying at infinity  for all $\bm w\in \mathcal{S}_{d-1}$ with $w_t>0).$ The function $\lambda_D(\bm w)$ satisfies a number of properties, including $\lambda_D(\bm w)\geq \max\{\bm w\}$ for all $\bm w\in \mathcal{S}_{d-1}.$ In the boundary case then $\lambda_D(\bm w)= \max\{\bm w\},$ and the variables in the random vector $\bm X$ exhibit AD. In addition, complete independence is achieved when $\lambda_D(\bm w)=1$ for all $\bm w\in \mathcal{S}_{d-1}.$ Furthermore, the coefficient $\eta_D$ can be obtained from the function $\lambda_D(\bm w)$ by setting $\bm w = \bm{1}_d\slash d$ where $\bm 1_d = (1, \ldots, 1)$ is a $d$-dimensional vector as then $\eta_D = [d\lambda_D(\bm{1}_d\slash d)]^{-1}.$ In a similar way to $\chi_D$ and $\eta_D,$ it can be shown by rearrangement of expression~\eqref{eq:lambdaddim} that $\lambda_D(\bm w)$ can be taken as the limit $\lambda_D(\bm w) = \lim_{r\to 1} \lambda_D(\bm w, r),$ where it exists, with
\begin{equation}\label{eq:lambdaddimr}
    \lambda_D(\bm w, r) = w_t \frac{\log(\Pr[F_{X_i}(X_i) > 1-(1-r)^{w_i\slash w_t}: \forall i \in D])}{\log(1-r)},\quad r\in(0,1), 
\end{equation}
for any $\bm w \in \mathcal{S}_{d-1}$ with $w_t>0$ for $t\in D.$

\subsection{Model definition and inference for copula} \label{subsection:modeldef}

Let us consider a $d$-dimensional random vector $\bm Y :=(Y_1, \ldots, Y_d) \in \mathbb{R}^d.$ We propose a dependence model for the copula of $\bm Y$ based on a mixture of multivariate Gaussian distributions with each of its margins being a mixture of Gaussians. We transform the margins of $\bm Y$ into uniform margins through $\bm U = T(\bm Y),$ where $T:\mathbb{R}^d\to [0,1]^d$ is applied componentwise, and then fit a copula to the random vector $\bm U,$ i.e., taking the dependence structure of $\bm Y.$ Working in a copula-based framework instead of considering the original scale is not novel \cite[e.g.,][]{Wadsworthetal2017}.

Assume now that we have a mixture of $k\geq 1$ components, where the $j^{th}$ component is a $d$-dimensional random variable $\bm Z_j:=(Z_j^1, \ldots, Z_j^d)$ where $j\in K=\{1, \ldots, k\}.$ Variables from different mixture components, i.e., $Z_j^i$ and $Z_{j'}^{i'}$ for any $j\neq j'\in K,$ are taken to be independent for all $i, \, i' \in D.$ Moreover, we assume that $\bm Z_j$ follows a multivariate Gaussian distribution, i.e., $\bm Z_j \sim \MVN({\bm \mu_j, \Sigma_j}),$ with mean vector $\bm\mu_j = (\mu_j^1, \ldots, \mu_j^d)'$ and variance-covariance matrix
\begin{equation}\label{eq:covmatrix}
	\Sigma_j = \begin{pmatrix}
		\sigma_{1j}^2 & \rho_j^{1,2}\sigma_{1j}\sigma_{2j} & \ldots &   \rho_j^{1,d}\sigma_{1j}\sigma_{dj} \\
		 \rho_j^{1,2}\sigma_{1j}\sigma_{2j}  & \sigma_{2j}^2 & \ldots   &  \rho_j^{2,d}\sigma_{2j}\sigma_{dj}  \\ 
		 \vdots & \vdots & \ddots & \vdots \\
		 \rho_j^{1,d}\sigma_{1j}\sigma_{dj} &  \rho_j^{2,d}\sigma_{2j}\sigma_{dj} & \ldots & \sigma_{dj}^2 
	\end{pmatrix},
\end{equation}
where $\rho_j^{m,n}\in [-1, 1]$ for $j \in K$ and $m\neq n\in D$ is the correlation between the $m^{th}$ and $n^{th}$ variables of the $\bm Z_j^{th}$ mixture component, and $\sigma_{ij}>0$ for all $j\in K$ and $i\in D.$ Consequently, we have that $Z_j^i\sim N(\mu_j^i, \sigma_{ij}^2)$ for $i \in D$ and $j \in K.$ As with any mixture model some conditions need to be imposed to ensure identifiability of the parameters of the mixture terms. Identifiability of our model is further complicated by the sole use of the copula structure, leading to other parameters of our model for $\bm Y$ not being identifiable on the copula scale. For these two reasons, respectively, we impose that $\mu_{j-1}^1 < \mu_j^1$ for $j=2, \ldots, k,$ and $(\bm \mu_1, \sigma_{11}^2) = (\bm 0_d, 1),$ where $\bm{0}_d=(0, \ldots, 0)$ is a $d$-dimensional vector of zeros. The latter condition ensures some referencing point when moving to the copula framework, as copulas are invariant to additive and scale transformations of $\bm Y$.

The random vector $\bm Y,$ whose distribution function is $F_{\bm Y}$ and $Y_i\sim F_{Y_i}$ $(i\in D),$ is then defined as below
\begin{equation}\label{eq:modelG}
    \bm Y=
       \bm Z_i \text{ with probability }p_i, \text{ for }i=1\ldots ,k.
\end{equation}
where $\bm Z_i =  \left(Z_i^1,Z_i^2, \ldots, Z_i^d\right)$, 
with $p_1\in (0, 1],$ $0\leq p_j\leq 1$ for $j = 2, \ldots k,$  and $\sum_{j=1}^k p_j=1.$ Furthermore, the survivor marginal distribution for random variable $Y_i$ $(i\in D)$ can be defined as
\begin{equation}\label{eq:marginalsurvivor}
    \overline{F}_{Y_i}(y) = \sum_{j=1}^k p_j\overline{\Phi}\left(\frac{y - \mu_j^i}{\sigma_{ij}}\right), \quad \text{for }y \in \mathbb{R},
\end{equation}
where $\overline{\Phi}$ denotes the standard Gaussian survivor distribution function and $k\geq 1.$ Similarly, the joint survivor distribution function of $\bm Y$ is given as
\begin{equation}\label{eq:jointsurvivor}
    \overline{F}_{\bm Y}(\bm y) = \sum_{j=1}^k p_j\Pr[Z_j^1 > y_1, \ldots, Z_j^d > y_d], \quad \text{for }\bm y=(y_1, \ldots, y_d)\in \mathbb{R}^d.
\end{equation}

In order to be able to work on a copula-based framework, we first need to transform the margins into uniform margins; this can be achieved via the probability integral transform. More specifically, if $T(\bm Y) = (F_{Y_1}(Y_1), \ldots, F_{Y_d}(Y_d))'$ with $Y_i\sim F_{Y_i}$ for $i \in D,$ then $\bm U = T(\bm Y)$ is a random vector with standard uniform margins. Therefore, we can fit the copula density of $\bm Y$ as follows
\begin{equation}\label{eq:copdens}
    c_{\bm Y}(\bm u; \bm \theta) = \frac{f_{\bm Y}\left(F_{Y_1}^{-1}(u_1), \ldots, F_{Y_d}^{-1}(u_d);\bm \theta\right)}{\displaystyle{\prod_{i=1}^d} f_{Y_i}\left(F_{Y_i}^{-1}(u_i)\right)}, \quad \bm u\in[0,1]^d,
\end{equation}
where $f_{Y_i}$ and $f_{\bm Y}$ are the density functions of $Y_i$ and $\bm Y,$ respectively, $F_{Y_i}^{-1}$ is the inverse cumulative distribution function (cdf) for $i\in D,$ and $\bm \theta=(\bm p, (\bm \mu_j, \bm \sigma_{\Sigma_j}, \bm \rho_{\Sigma_j}): j\in K)$ \linebreak is the vector of model parameters, where $\bm p=(p_1, \ldots, p_{k-1}),$ $\bm \mu_j=(\mu_{1}^j, \ldots, \mu_{d}^j),$ \linebreak $\bm \sigma_{\Sigma_j} = (\sigma_{1j}, \ldots, \sigma_{dj})$ and $\bm \rho_{\Sigma_j} = (\rho_j^{1,2}, \ldots, \rho_j^{d-1, d})$ from the variance-covariance matrix $\Sigma_j$. This results in $k\left(1+d(d-3)\slash 2\right)-d-2$ model parameters that need to be estimated, which increases linearly with $k,$ but quadratically with dimension $d.$ 

When the margins of a random variable $\bm X\in \mathbb{R}^d$ are unknown, which is often the case, then $\bm U\in [0,1]^d,$ with uniform $[0,1]$ margins, is obtained through componentwise rank transform of the data $\bm X.$ For $n$ independent and identically distribution (i.i.d.) observations from $\bm X,$ which are assumed to have copula family $c_{\bm Y}(\bm u; \bm \theta)$, inference on model~\eqref{eq:copdens} is performed by maximum likelihood estimation (MLE) of the vector of parameters $\bm \theta$ with the log-likelihood function
\begin{equation}\label{eq:loglike}
    \ell(\bm \theta) = \sum_{t=1}^n c_{\bm Y}(\bm u_t;\bm\theta), 
\end{equation}
where $\bm u_t\in [0,1]^d, \, t = 1,\ldots, n, \, n\geq 2$ are the transformed sample of the $\bm X$ variables on uniform margins. The identifiability constraints on the parameters are imposed within the log-likeli\-hood function; see Section~\ref{supsec:casestudy} of the Supplementary Material for more details.

To aid the inference procedure, information about the proposed graphical structure of $\bm Z_j$ $(j\in K)$ components can be exploited, enabling both expert judgement coupled with a reduction in the dimensionality of the parameters space of our copula model. For instance, if it is considered appropriate to model variables $Z_j^m$ and $Z_j^n$ for $m\neq n \in D$ for some $j\in K$ as conditionally independent given the remaining $d-2$ variables from the mixture component $\bm Z_j$, this information can be embedded in the likelihood function. Alternatively, fewer parameters need to be estimated if we have pairwise exchangeability, that is, for each mixture component $\bm Z_j,$ all means and variances are assumed to be identical, i.e., $\mu_j^1 = \ldots = \mu_j^d$ and $\sigma_{1j}^2 = \ldots = \sigma_{dj}^2$ for each $j\in K;$ under this assumption, only a single mean and a single variance are inferred for each one of the $k$ mixture components, rather than $2d$ parameters per component. We note, however, that we are still required to estimate the $d(d-1)\slash 2$ correlation parameters.

\subsection{Extremal dependence properties} \label{subsection:extremaldep}

The sub-asymptotic measure $\chi_D(r),$ given by expression~\eqref{eq:chiddim}, can be found numerically for any vector of model parameters $\bm \theta.$ Although it is known that for a Gaussian copula, $\chi_D(r)\to 0$ and $\eta_D(r) \to (\bm{1}_d'\Sigma_{\bm \rho}^{-1}\bm{1}_d)^{-1},$ as $r\to 1,$ where $\Sigma_{\bm \rho}\in \mathbb{R}^{d\times d}$ is the underlying Gaussian correlation matrix with $d\geq 2$ \citep{Joe2014}, we will show that $\chi_D(r)$ can be arbitrarily close to 1 for any $r\in (0,1)$ with the Gaussian mixture copula, and thus being able to capture key features of the data at sub-asymptotic levels, even for AD variables. In particular, to illustrate this property we assume that $\bm \mu_j = \bm 0_d$ for $j=1, \ldots, k-1$ and $\bm \mu_k=\mu\bm{1}_d$ with $\mu >0,$ and $\bm\sigma_{\Sigma_j}=\bm{1}_d$ for all $j\in K.$ Additionally, consider $\rho_j^{m,n}=\rho$ for all $m\neq n \in D$ and $j\in K,$ and let $p_j = (k-p)\slash(k(k-1))$ for $j = 1, \ldots, k-1$ and $p_k = p\slash k$ for $0\leq p\leq 1.$ This structure allows us to represent our model with $k-1$ similar mixture terms and with one different mixture term. Furthermore, less weight is assigned to the $k^{th}$ mixture component given that $0 \leq p_k \leq 1\slash k.$ While these are rather simplistic assumptions, the same arguments hold if we take $\bm \mu_j = j\varepsilon$ for small  $\varepsilon>0$ (satisfying in this way the ordering condition) for example, therefore ensuring an identifiable model.

Given this structure, $Y_i$ $(i\in D)$ are identically distributed and its survivor marginal distribution, given in expression~\eqref{eq:marginalsurvivor}, simplifies to
\begin{equation} \label{eq:jointsurvsimpl}
    \overline{F}_{Y_i}(y) = \left(1-\frac{p}{k}\right)\overline{\Phi}(y) + \frac{p}{k}\overline{\Phi}(y-\mu), \quad y\in \mathbb{R}.
\end{equation}
Similarly, we can define the joint survivor distribution of $\bm Y,$ given in expression~\eqref{eq:jointsurvivor}, on the diagonal as
\begin{equation} \label{eq:jointsurvsimpla}
    \overline{F}_{\bm Y}(y\bm{1}_d) = \left(1-\frac{p}{k}\right)\overline{\Phi}_d(y\bm{1}_d; \Sigma_\rho) + \frac{p}{k}\overline{\Phi}_d((y-\mu)\bm{1}_d; \Sigma_\rho), 
\end{equation}
where $\overline{\Phi}_d$ is the standard multivariate Gaussian survivor distribution function with correlation matrix $\Sigma_\rho,$ i.e., with all off-diagonal entries $\rho,$ and $d\geq 2.$ 

Consider now having $\mu>0$ to be large. As $y\to \infty,$ we have that
\begin{equation}\label{eq:overlinef}
    \overline{F}_{Y_i}(y) \sim \frac{p}{k}\overline{\Phi}(y-\mu) \quad \text{and} \quad \overline{F}_{\bm Y}(y\bm{1}_d) \sim\frac{p}{k}\overline{\Phi}_d((y-\mu)\bm{1}_d; \Sigma_\rho).
\end{equation}
Recall that $\chi_D(r),$ defined in expression~\eqref{eq:chiddim}, is in terms of standard uniform variables. So that we are able to determine $\chi_D(r)$ under the imposed conditions, we need to express it in terms of the variables $Y_i$ for all $i\in D.$ Owing to the assumptions made on the parameters, we have common margins; hence we let $y=F_{Y_i}^{-1}(r)$ for $i \in D.$ Expression~\eqref{eq:chiddim} can then be rewritten as
\begin{equation}\label{eq:chiddimN}
    \chi_D(r)=\frac{\Pr[Y_i > F^{-1}_{Y_i}(r): \forall i \in D]}{\Pr[Y_1 > F^{-1}_{Y_1}(r)]}, \quad \text{for }r\in (0, 1). 
\end{equation}

With the imposed conditions on the mean, variance and correlation parameters, it follows that the sub-asymptotic extremal dependence measure $\chi_D(r),$ can be explicitly written~as
\begin{equation}\label{eq:condprob}
    \chi_D(r) \sim \frac{\overline{\Phi}_d\left((F_{Y_i}^{-1}(r)-\mu)\bm{1}_d; \Sigma_\rho\right)}{\overline{\Phi}\left(F_{Y_i}^{-1}(r)-\mu\right)}, \quad \text{as }r\to 1.
\end{equation}

Now consider letting $\mu\to \infty$ and $p\to 0$ as $r\to 1,$ with $\overline{\Phi}(\mu)\slash p \to 0$ also as $r\to 1.$ This ensures that the marginal tail of the distribution of $Y_i$ $(i \in D)$ is dominated by the $k^{th}$ mixture component. Specifically, we have that $1-r = \overline{F}_{Y_i}(y),$ so from expression~\eqref{eq:jointsurvsimpl}, when $\mu = y,$
\begin{equation}
    1-r = \frac{p}{k}\overline{\Phi}(0) + \left(1-\frac{p}{k}\right)\overline{\Phi}(\mu) = \frac{p}{2k} + \left(1-\frac{p}{k}\right)\overline{\Phi}(\mu),
\end{equation}
since $\overline{\Phi}(0) = 1\slash 2.$ As $p\to 0$ and $\mu \to \infty,$ with $\overline{\Phi}(\mu)\slash p \to 0$ as $r\to 1,$ we then have that
\begin{equation}
    1-r = \frac{p}{2k} + \mathcal{O}(\overline{\Phi}(\mu)) = \frac{p}{2k}\left(1+\mathcal{O}\left(\frac{\overline{\Phi}(\mu)}{p}\right)\right) = \frac{p}{2k}\left(1 + o(1)\right).
\end{equation}
Thus, $1-r\sim p\slash (2k)$ as $r\to 1,$ and $F_{Y_i}^{-1}(r)\sim \mu$ as $r\to1.$ It then follows from expression~\eqref{eq:condprob} that
\begin{equation}\label{eq:condprobsim}
    \chi_D(r) \sim \frac{\overline{\Phi}_d\left(\bm{0}_d; \Sigma_\rho\right)}{\overline{\Phi}(0)}=2\overline{\Phi}_d\left(\bm{0}_d; \Sigma_\rho\right).
\end{equation}
Thus, we have that
\begin{equation}
    \chi_D(r) \to 2\overline{\Phi}_d\left(\bm{0}_d; \Sigma_\rho\right), \quad \text{as } r\to 1.
\end{equation}
Given that $\overline{\Phi}_d(\bm{0}_d; \Sigma_0)= (1\slash 2)^d$ and $\overline{\Phi}_d(\bm{0}_d; \Sigma_1)= 1\slash 2,$ by suitable changes in $\rho,$ the measure $\chi_D(r)$ can exceed any arbitrary level up to 1. This is possible when the mode $\mu$ is sufficiently larger than the other mixture modes, and the $k^{th}$ mixture probability $p_k$ approaches 0. Relaxing some or all of these constraints will only allow for more general and richer joint behaviour. We note that similar results can be derived for measures $\eta_D(r)$ and $\lambda_D(\bm w, r)$ from expressions~\eqref{eq:etarddim} and \eqref{eq:lambdaddimr}, respectively.

\section{Model fit and diagnostics} \label{section:modelfit}

\subsection{Overview} \label{subsubsection:overview}

The fit of the proposed model~\eqref{eq:copdens} is assessed in a range of data sets exhibiting different dependence structures; we show several cases here and refer the reader to Section~\ref{supsubsec:modelfit} of the Supplementary Material for additional cases. For the model selection procedure, we use the Akaike information criterion \cite[AIC;][]{Akaike1974}. We compare estimates of the extremal dependence measures $\chi_D(r)$ and $\eta_D(r)$ from expressions~\eqref{eq:chiddim} and \eqref{eq:etarddim}, respectively, obtained through the model fit with their empirical counterparts and their true values. The marginal and joint empirical probabilities needed to estimate the numerator and denominator of the empirical measures in expressions~\eqref{eq:chiddim} and \eqref{eq:etaddim} are computed by considering the proportion of points lying in the regions $(u_1, 1)$ and $(u_1, 1)\times \ldots\times (u_d, 1),$ respectively. Furthermore, pointwise $95\%$ confidence intervals are obtained for both measures by computing the empirical $\chi_D(r)$ and $\eta_D(r)$ for $n$ bootstrap samples of the data.

When the interest is in regions where at least one variable is extreme, we compare probabilities obtained with the model fit and their empirical values in regions of the form $(u_1, 1)\times (0, u_2) \times \ldots \times (0, u_d)$ when considering $U_1$ being extreme (and hence $u_1$ is close to 1), for example. Although this relates with function $\lambda_D(\bm w, r)$ given in expression~\eqref{eq:lambdaddimr}, we instead obtain such probabilities by considering extremal regions $A_{\bm w},$ for $\bm w\in \mathcal{S}_{d-1},$ with this region defined by standard exponentially distributed variables. More specifically, in a bivariate setting we have that
\begin{equation}
    A_{w}= \left\{X_1^E > \max\left\{\frac{w}{1-w}, 1\right\}u^E, X_2^E > \max\left\{\frac{1-w}{w}, 1\right\}u^E\right\}
\end{equation}
where $(X_1^E, X_2^E)$ is a $2$-dimensional random vector with standard exponential random variables $X_i^E$ for $i=1,2,$ and $u^E$ is a threshold level for $\max\{X_1^E, X_2^E\};$ see Section~\ref{supsec:appendixpaper2} of the Supplementary Material for more details. When moving to a $d$-dimensional setting, we consider the probability $\Pr((X_1^E, \ldots, X_d^E) \in A_{\bm w} \mid \max_{i \in D} \{X_i^E\} > u^E)$ as a function of $\bm w.$ 

We consider a range of copula families that exhibit different dependence structures to assess the performance of the Gaussian mixture copula. More specifically, for the case where the underlying data are AI, we consider an inverted extreme value copula with logistic dependence structure \citep{LedfordTawn1997}, since this copula is known to have $\chi_D=0.$ Following the same reasoning, we consider an extreme value copula with logistic dependence structure \citep{Gumbel1960} to assess the fit given by our model when the data are AD, since this copula has $\chi_D>0.$ To show the performance of the Gaussian mixture copula with non-exchangeable underlying data (i.e., data showing asymmetries), an extreme value copula with an asymmetric logistic dependence structure \citep{Tawn1988} is used. Finally, we assess the fit of our copula model with more complex type data by considering a particular specification of the weighted copula model (henceforth referred to as WCM) proposed by \citet{Andreetal2024}. In all cases, the non-exchangeable Gaussian mixture copula model is used. However, in the AI and AD cases, the performance of the model may improve if the exchangeable model is used instead. The results for the cases where the underlying data are AI and from a WCM specification are given in Section~\ref{supsubsec:aidata} and \ref{supsubsec:wcm}, respectively, of the Supplementary Material, and show a good fit of the Gaussian mixture copula for both cases. Specifically, when the underlying data is AI, a good representation of the joint extremal behaviour is obtained with the models with $k=1, 2$ and $k=3$ mixture components. For the case where the data is generated from a WCM, a significant improvement in fit is observed when considering the $k=2$ model, compared to the model with a single component.

\subsection{Asymptotically dependent data} \label{subsubsec:ad}

\begin{figure}[h!]
    \centering
    \includegraphics[width=0.9\textwidth]{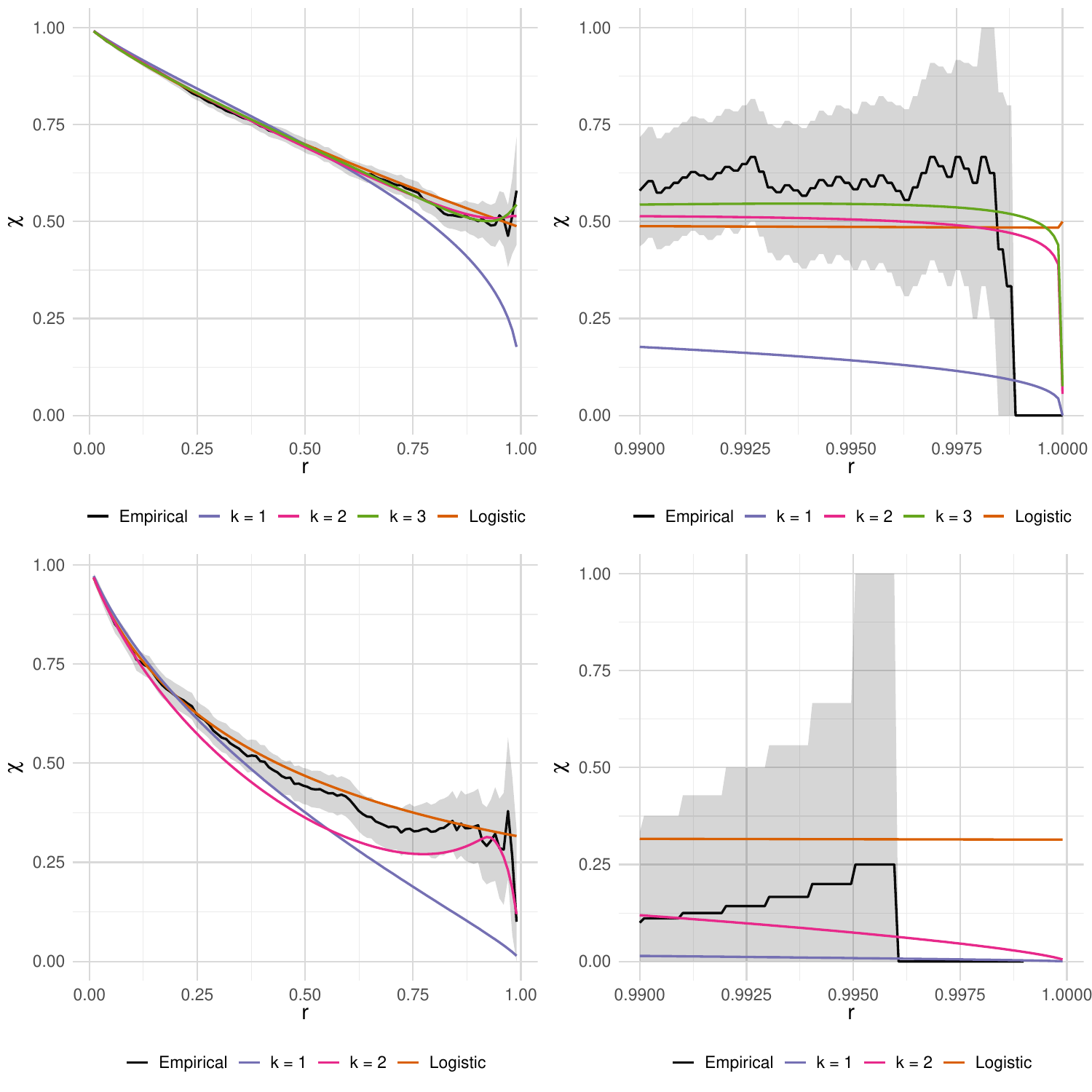}
    \caption{Estimates of $\chi_D(r)$ for $r\in (0.1)$ with true (in orange) and empirical (in black) values also shown. These are zoomed in for $r\in[0.99,1)$ on the right. The pointwise $95\%$ confidence intervals for the empirical $\chi_D(r)$ are obtained through bootstrap. When $d=2$ (top), models with $k = 1,2$ and $3$ mixture components are considered, whereas when $d=5$ (bottom) models with only $k=1$ and $2$ mixture components are studied.}
    \label{fig:ad}
\end{figure}

The performance of the Gaussian mixture copula is assessed on a AD copula, specifically the extreme value copula with logistic dependence structure. When $d = 2,$ data are generated with dependence parameter $\alpha_L= 0.6,$ and sample size $n = 5000.$ As mentioned previously, this model exhibits AD with $\chi_2 = 2-2^{\alpha_L}.$ We consider Gaussian mixture copulas with $k=1, 2$ and $3$ mixture components. The decrease in AIC with $k>1$ relative to when $k=1$ is $-219.28$ for $k=2$ and $-226.18$ for $k=3,$ indicating that the copula with $k=3$ mixture components is the one that best fits the data. This is further supported when comparing measure $\chi_2(r)$ for $r\in (0,1)$ obtained with these three model fits. The results are shown in the top panel of Figure~\ref{fig:ad}, where in the right $\chi_2(r)$ is zoomed in for $r\in [0.99, 1)$. We can see that in the case where $k=1,$ the fitted model is AI, thus clearly under-estimates $\chi_2(r)$ for $r>0.6,$ with the bias increasing as $r\to 1.$ On the other hand, the Gaussian mixture copulas with $k=2$ and $3$ are able to account for the behaviour of the joint tail at levels $r$ very close to 1 with values of $\chi_2(r)$ close to the true value over this region. This approximate finding of AD is consistent with the underlying data which is known to exhibit AD. The corresponding plot for $\eta_2(r)$ is given in the left panel of Figure~\ref{fig:adeta} of the Supplementary Material, showing similar findings. A similar study with a smaller sample size $(n=1000)$ is presented in Figure~\ref{fig:adsup} of the Supplementary Material. It can be seen that now the model with $k=2$ components is not able to capture the extremal behaviour, i.e., at levels of $r$ close to 1. Higher sample sizes may improve the flexibility of this model specification and its ability to capture $\chi_2(r)$ at levels of $r$ very close to 1, as shown by the case when $n=5000.$

Consider now a $d=5$ dimensional setting with $n=1000.$ Due to the larger number of parameters we study only $k=1$ and $2$ mixture components. When considering $k=2$ components, a mixing probability estimate of $\hat p_1 = 0.98$ is obtained. Despite $\hat p_1$ being so close to one, the extra component adds more flexibility to the modelling of the data, resulting in a decrease in AIC of $-148.69$ in relation to $k=1.$ This is also visible in the bottom right panel of Figure~\ref{fig:ad}, where, as before, $\chi_5(r)$ is zoomed in for $r\in[0.99,1).$ Although the results suggest that for this setting we probably need $k>2,$ and a larger $n,$ to get a better estimate of $\chi_5(r)$ from the Gaussian mixture copula, the model with $k=2$ components is able to capture the joint tail behaviour well for levels $r\in (0.9, 0.99],$ despite it under-estimating $\chi_5(r)$ for lower values of $r.$ The results for $\eta_D(r)$ are given in the right panel of Figure~\ref{fig:adeta} of the Supplementary Material, and show similar findings. The empirical estimates and their pointwise confidence intervals are 0 for $r>0.998$ and $r>0.996$ for $d=2$ and $d=5,$ respectively, since there are no observations that jointly exceed such values. Thus, the empirical $\chi_D(r)$ fails to characterise the joint behaviour beyond these values of $r.$ As shown by the right panel, this is not the case for the Gaussian mixture copula, particularly for $d=2,$ as the estimates of $\chi_2(r)$ for the $k=2$ and $k=3$ models lie close to the truth for $r$ very close to 1.

\subsection{Non-exchangeable data} \label{subsubsec:asy}

\begin{figure}[h!]
    \centering
    \includegraphics[width=0.9\textwidth]{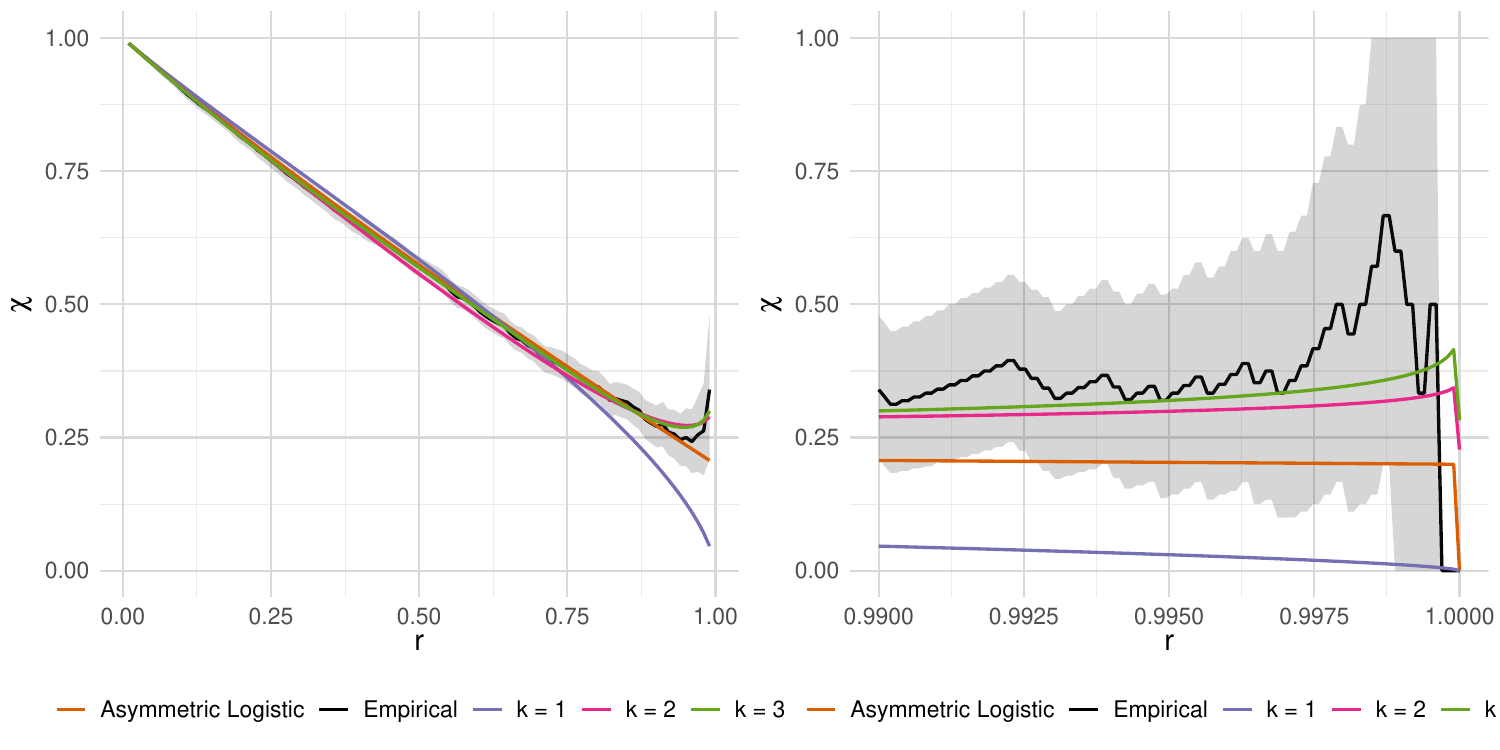}
    \caption{Estimates of $\chi_2(r)$ for $r\in (0.1)$ with true (in orange) and empirical (in black) values also shown. These are zoomed in for $r\in[0.99, 1)$ on the right. The pointwise $95\%$ confidence intervals for the empirical $\chi_2(r)$ are obtained through bootstrap.}
    \label{fig:asy}
\end{figure}

We generate $n=5000$ samples from a bivariate extreme value copula with asymmetric logistic dependence structure with dependence parameter $\alpha_{A} = 0.2$ and asymmetry parameters $t_1=0.2$ and $t_2=0.8$ This copula has $\chi_2 = t_1+t_2-\left(t_1^{1\slash \alpha_A} + t_2^{1\slash \alpha_A}\right)^{\alpha_A}.$ As with the previous case, we consider the Gaussian mixture copula with $k=1, 2, 3.$ From the results shown in Figure~\ref{fig:asy}, we see that the $k=1$ model is not able to capture the extremal behaviour of the data, while the $k=2$ and $k=3$ models provide a good fit overall for $\chi_2(r)$ up to $r$ very close to 1. This is in agreement with the AIC values, where a decrease of $-978.76$ for $k=2$ and of $-1020.02$ for $k=3$ relatively to the $k=1$ model is observed, so there is not much difference in AIC for $k=2$ or $k=3.$ Given that the $k=2$ and $k=3$ models seem to capture the joint behaviour for all $r,$ it is sufficient to consider a simpler model with $k=2$ in this particular case. The results for $\eta_2(r),$ given in Figure~\ref{fig:asysup} of the Supplementary Material, show similar conclusions.

\begin{figure}[h!]
    \centering
    \includegraphics[width=0.9\textwidth]{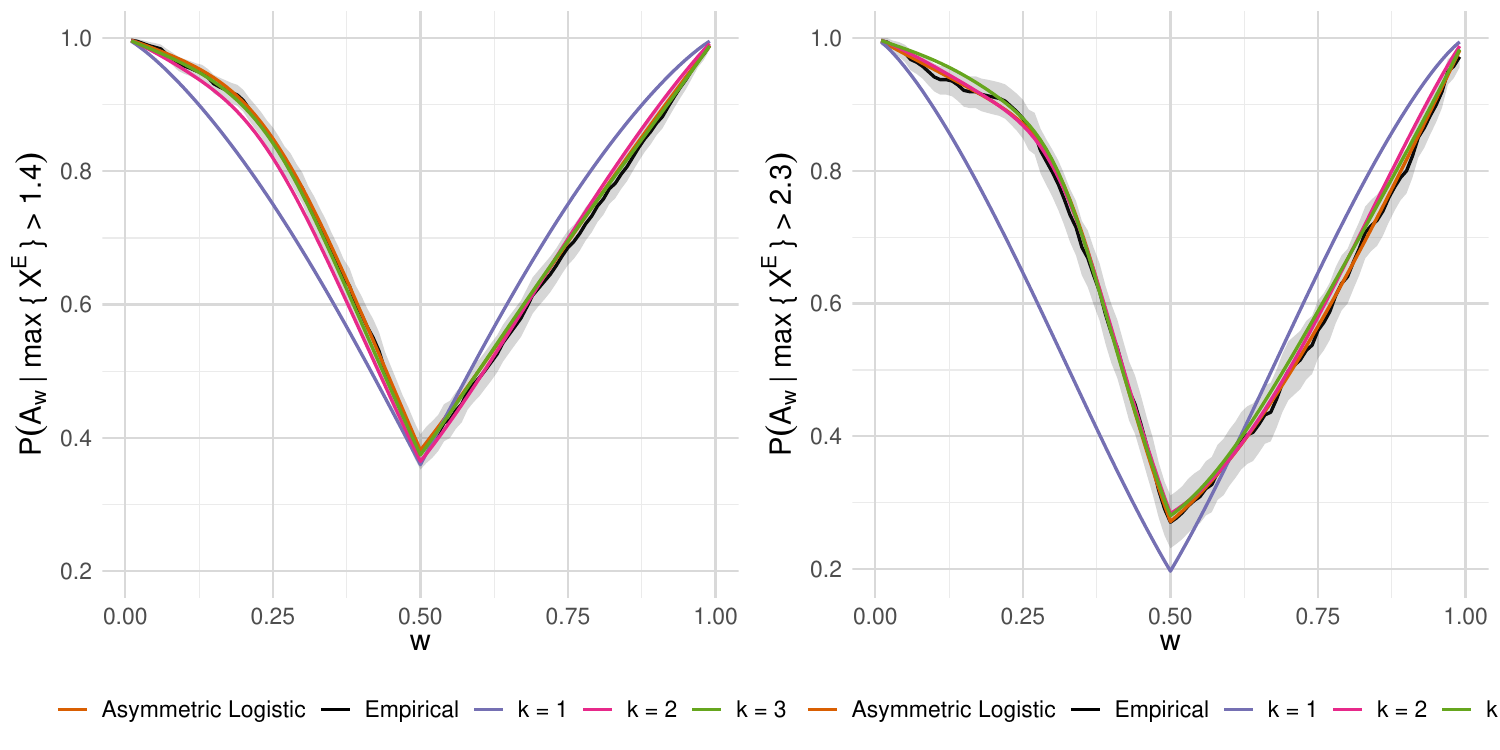}
    \caption[Comparison between the estimates of probabilities for two large values $u^E=\{1.4, 2.3\}$ with true (in orange) and empirical (in black) values also shown. The pointwise $95\%$ confidence intervals for the empirical probabilities are obtained through bootstrap.]{Comparison between the estimates of probabilities $\Pr(A_w\mid \max_{i=1,2}\{X_i^E\}> u^E)$ for two large values $u^E=\{1.4, 2.3\}$ with true (in orange) and empirical (in black) values also shown. The pointwise $95\%$ confidence intervals for the empirical probabilities are obtained through bootstrap.}
    \label{fig:asybw}
\end{figure}

Further, we assess the performance of the Gaussian mixture copula along different rays $w \in \mathcal{S}_1$ and compute the probability $\Pr(A_w\mid \max_{i=1,2}\{X_i^E\}> u^E).$ The results for the $0.75$ and $0.90$ quantiles $u^E = \{1.4, 2.3\}$ of $\max_{i=1,2}\{X_i^E\},$ respectively, are shown in Figure~\ref{fig:asybw} in the left and right panels, respectively. Similarly to measures $\chi_2(r)$ and $\eta_2(r),$ the $k=2$ and $k=3$ models capture the extremal behaviour at all $w$ considered for either $u^E.$ In particular, they lie within the pointwise $95\%$ confidence intervals for the empirical probabilities. On the other hand, the Gaussian mixture copula with $k=1$ under-estimates the joint behaviour for $w\leq 0.5,$ and over-estimates otherwise, lying outside of the pointwise $95\%$ confidence intervals for the most $w$. This is particularly pronounced for higher $u^E$, as shown by the right panel. 

\section{Case study: air pollution data} \label{section:casestudy}

\subsection{Data description and previous analysis}

We apply the Gaussian mixture copula to the $5$-dimensional seasonal air pollution data set analysed by \citet{HeffernanTawn2004}, which consider the joint behaviour of random variables conditionally on one of them being large. Contrary to the Gaussian mixture copula, the conditional approach requires the definition of an extremal region of the form $\{X_2, \ldots, X_d\} \mid \{X_1 > u\}$ for some large marginal threshold $u,$ for example. In their study, \citet{HeffernanTawn2004} take $u$ to be the $0.9$ marginal quantile. However, as stated by \citet{LiuTawn2014}, this conditional approach is not self-consistent as considering different conditioning variables, i.e., given $\{X_j>u\}$ not given $\{X_i>u\}$ for $j=2, \ldots, d,$ may lead to different conclusions in the joint region $\{X_i>u, X_j> u\}$ $(i \in D)$, which is not the case for the Gaussian mixture copula model. 

The data set includes daily maxima of the hourly means of ground level measurements of ozone $(O_3),$ nitrogen dioxide $(NO_2),$ nitrogen oxide $(NO),$ sulphur dioxide $(SO_2)$ and particulate matter $(PM_{10})$ recorded at Leeds, UK, from 1994 to 1998. In order to remove the temporal non-stationarity, \citet{HeffernanTawn2004} divide the data set into two seasons, winter from the months of November to February, and summer from the months of April to July.  In their analysis, the pairs $(NO_2, NO),$ $(NO, PM_{10})$ and $(NO_2, PM_{10})$ were judged to exhibit AD in the winter season, with the remaining pairs (in both seasons) indicating the presence of AI. In our analysis, we denote the variables after rank transformation to uniform $(0,1)$ variables as $O_3^*,$ $NO^*_2,$ $NO^*,$ $SO^*_2$ and $PM^*_{10}.$ 

\subsection{Pairwise analysis} \label{subsection:pairwise}

\begin{figure}[h!]
    \centering
    \includegraphics[width=\textwidth]{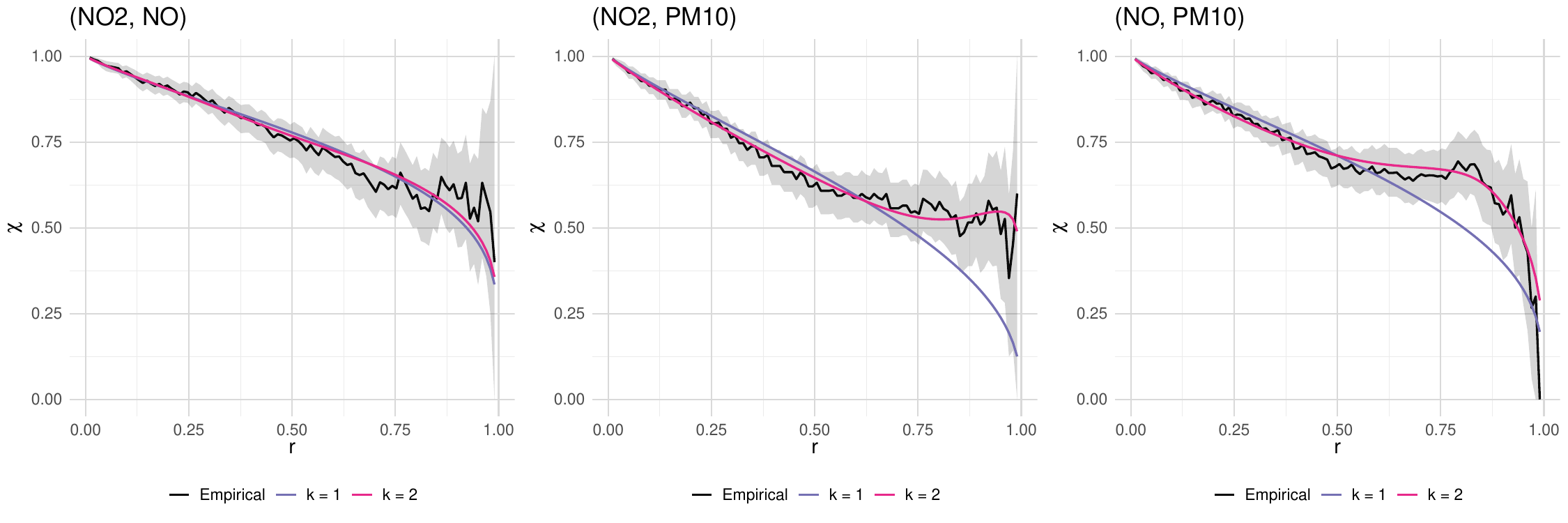}
    \caption{Estimates of $\chi_2(r)$ for $r\in (0.1)$ with empirical (in black) values also shown for pairs $(NO_2, NO)$ (left), $(NO_2, PM_{10})$ (middle) and $(NO, PM_{10})$ (right). The pointwise $95\%$ confidence intervals for the empirical $\chi_2(r)$ are obtained through bootstrap.}
    \label{fig:pairwise}
\end{figure}

 We apply our Gaussian mixture copula with $k=1$ and $2$ mixture components to the three pairs that \citet{HeffernanTawn2004} identified as being potentially AD to determine whether we obtain similar results. The change observed in the AIC values shown in Table~\ref{tab:pairwise} (denoted by $\text{AIC}_{k_1-k_2}$) suggest that $k=2$ is the most suitable model for all pairs except $(NO, NO_2),$ for which there is a small increase in AIC for $k=2$ when compared to the $k=1$ model. These results are in agreement with the model-based $\chi_2(r)$ obtained for the three pairs for $r\in (0,1),$ as shown in Figure~\ref{fig:pairwise}. In particular, the estimated $\chi_2(r)$ given by the mixture model with $k=2$ closely aligns with the behaviour of the empirical measure across all $r\in (0.1).$ On the other hand, it is clear that the $k=1$ model is not able to capture the asymptotic behaviour of pairs $(NO_2, PM_{10})$ and $(NO, PM_{10}),$ as it under-estimates the empirical $\chi_2(r)$ for $r>0.6$ by approaching 0 quicker. However, it appears sufficient for pair $(NO_2, NO).$ Although the estimated mixing probabilities are far from 0 or 1, the AIC and $\chi_2(r)$ results for pair $(NO_2, NO)$ indicate that adding an extra component is not necessary, as little to no difference is notable in the considered diagnostics. Furthermore, for pairs $(NO, NO_2)$ and $(NO_2, PM_{10}),$ the empirical $\chi_2(r)$ is clearly positive, which is also mirrored by the sub-asymptotic model-based $\chi_2(r)$ obtained by the $k=1$ and $k=2$ models for pair $(NO, NO_2),$ and by the $k=2$ for pair $(NO_2, PM_{10}).$ These results agree with the findings of \citet{HeffernanTawn2004}. Lastly, the results for $\eta_2(r),$ given in Figure~\ref{fig:pairwisesup} of the Supplementary Material, lead similar conclusions. For pair $(NO, PM_{10}),$ the estimated $\chi_2(r)$ from the $k=2$ model approaches 0 as $r\to 1,$ suggesting AI. In this case, measure $\eta_2(r)$ provides more insight, with $\eta_2(r)\to 0.75$ as $r\to 1,$ meaning that the extremes of $(NO, PM_{10})$ exhibit positive dependence. 

\begin{table}[h!]
    \centering
    \caption{Change in AIC values obtained for the Gaussian mixture copula for $k=2$ relative to when $k=1$ for pairs $(NO_2, NO)$, $(NO_2, PM_{10})$ and $(NO, PM_{10})$. The estimated mixing probabilities $(\hat p_1, \hat p_2)$ are reported for the $k=2$ model. All the values are rounded to 2 decimal places.}
    \begin{tabular}{l|c|c}
         Pair &  $\text{AIC}_{k_1-k_2}$ & $(\hat p_1, \hat p_2)$\\
         \hline
        $(NO_2, NO)$ & $4.01$ & $(0.37, \, 0.63)$  \\
        $(NO_2, PM_{10})$ & $-34.40$ & $(0.91, \, 0.09)$ \\
        $(NO, PM_{10})$ & $-50.87$ & $(0.78, \, 0.22)$ \\
        \hline
    \end{tabular}
    \label{tab:pairwise}
\end{table}


\subsection{Trivariate analysis}  \label{subsection:trivariate}

\begin{figure}[h!]
    \centering  
    \includegraphics[width=0.5\textwidth]{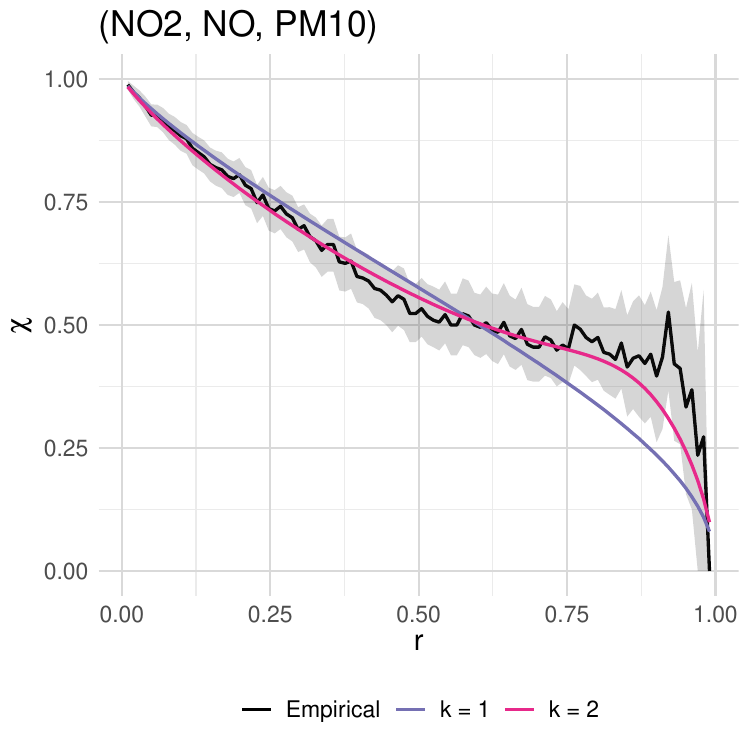}
    \caption{Estimates of $\chi_3(r)$ for $r\in (0.1)$ with empirical (in black) values also shown for the triple $(NO_2, NO, PM_{10})$. The pointwise $95\%$ confidence intervals for the empirical $\chi_3(r)$ are obtained through bootstrap.}
    \label{fig:appd3}
\end{figure}

Before analysing the full data set, we apply the Gaussian mixture copula with $k=1$ and $2$ mixture components to the triple consisting of the pollutants studied in the bivariate setting, i.e., $(NO_2, NO, PM_{10}),$ in the winter season to assess if the triple provides evidence for AD. Even if each of these pairs were AD, the triple being AD does not necessarily follow. However, if one pair (e.g., $(NO, PM_{10})$) were AI, then the triple must also be AI. The decrease in AIC for $k=2$ relative to when $k=1$ is of $-61.22,$ meaning that the $k=2$ provides the best fit to the triple according to this criterion. In addition, the mixing probabilities obtained for the $k=2$ model are $(\hat p_1, \hat p_2) = (0.73, 0.27),$ indicating that an extra Gaussian component allows for a more flexible fit. This can also be seen with the $\chi_3(r)$ estimates given in Figure~\ref{fig:appd3}. Whilst the true $\chi_3(r)$ is unknown, when comparing the model-based estimates with the empirical values, the $k=2$ model is able to capture the joint behaviour for all $r \in (0,1).$ The same is not true with $k=1,$ as it appears to over-estimate the empirical $\chi_3(r)$ for smaller $r$ and clearly under-estimate $\chi_3(r)$ for $r>0.75.$ Given that pair $(NO, PM_{10})$ exhibits AI according to the pairwise analysis, it is not a surprise that both $k=1$ and $k=2$ indicate that $NO_2,$ $NO$ and $PM_{10}$ cannot all be extreme at the same time, which is consistent with our findings from the three pairwise analysis. The results for $\eta_3(r)$ are shown in Figure~\ref{fig:appd3sup} of the Supplementary Material, for which the same conclusions can be drawn. Similarly to the pair $(NO, PM_{10}),$ the extremes of the triple $(NO_2, NO, PM_{10})$ are jointly positively dependent as $\eta_3(r) \to 0.62$ as $r\to 1$ for both $k=1$ and $k=2$ models.

\begin{figure}[h!]
    \centering
    \includegraphics[width=0.9\textwidth]{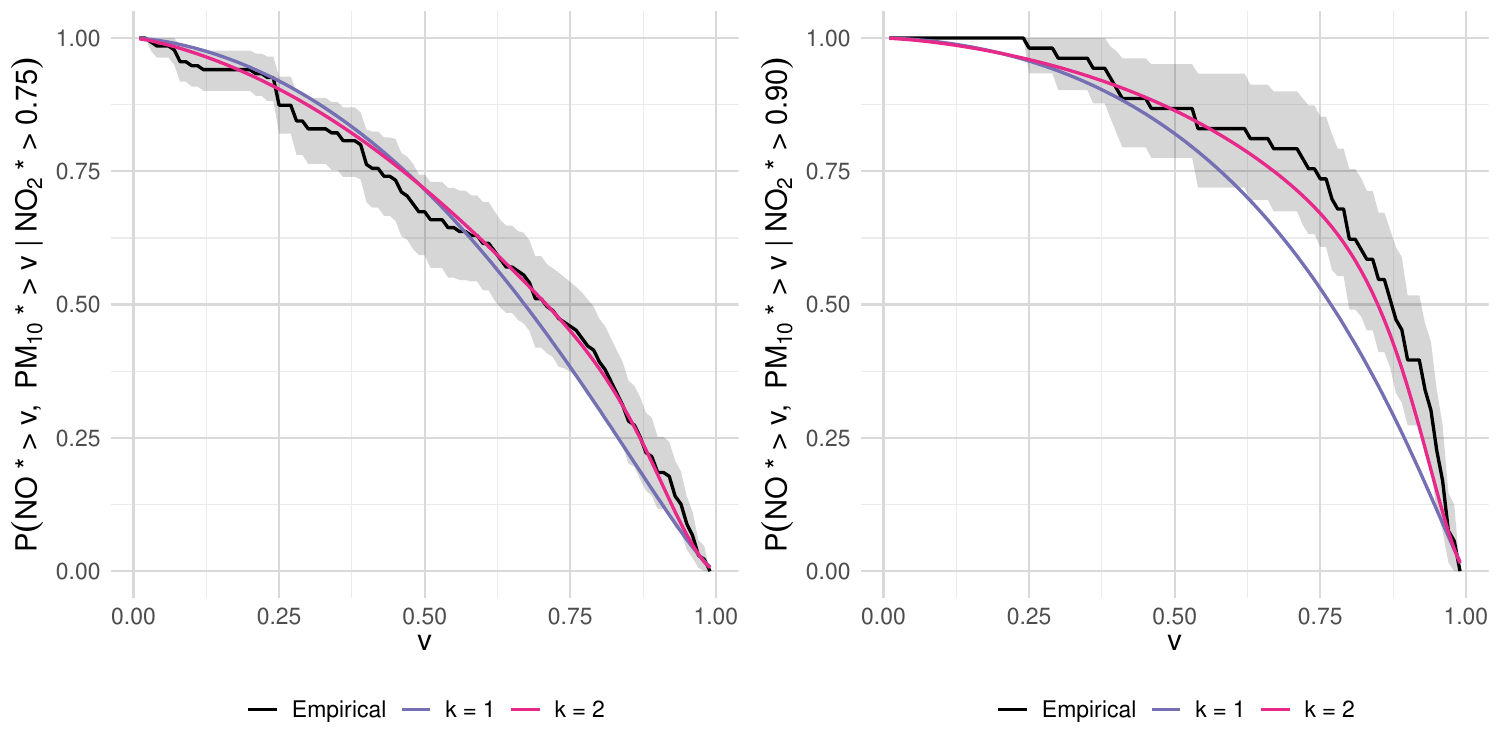}
    \caption{Comparison between model-based probabilities $\Pr(NO^* > v, PM^*_{10} > v \mid NO^*_2 > u)$ for two large values $u=\{0.75, 0.90\}$ given by the Gaussian mixture copula with $k=1$ (in purple) and $k=2$ (in pink) components. The empirical probability is given in black, and its pointwise 95$\%$ confidence intervals are obtained through bootstrap.}
    \label{fig:probd3}
\end{figure}

We further assess the performance of the Gaussian mixture copula by considering the behaviour of the remaining variables when conditioning on one variable being large. More specifically, we are interested in probabilities where at least one variable is extreme, e.g., of the form $\Pr(NO^* > v, PM^*_{10} > v \mid NO_2^* > u)$ with $v \in (0, 1)$ and some large $u.$ Considering such probabilities are key to learn about the risk of one pollutant, in this case $NO_2,$ exceeding a large level, as well as its impact on other pollutants, whether they too exceed or not a high level. Similarly to the measure $\chi_3(r)$, we compare the probabilities for both model fits with their empirical counterpart for $u=\{0.75, 0.90\};$ the results are shown in Figure~\ref{fig:probd3}. There is a clear the difference between the $k=1$ and $k=2$ models, with an improvement shown by $k=2$ when $u=0.9.$ In particular, the probabilities across all $v\in (0,1)$ lie within the empirical pointwise $95\%$ confidence intervals for both $u.$ The same is not true for the $k=1$ model when $u=0.90,$ suggesting that the $k=1$ model may perform poorly when at least one variable exceeds a very high level, such as 0.90. We note that for $v\leq 0.25$ and $u = 0.9,$ the lower and upper bounds of the confidence intervals for the empirical probability coincide and are equal to 1.

Further conclusions about the dependence between the variables can be drawn by exploring the graphical structure of the fit provided by each model. To do so, we analyse the precision matrices estimated from the $k=1$ and $k=2$ models, denoted by $\Sigma^{-1}_{\bm \rho\, (k=1)}$ and $\Sigma^{-1}_{\bm \rho, j\, (k=2)}$ for $j=1,2,$ respectively; their off-diagonal values are given in Table~\ref{tab:precd3}. From $\Sigma_{\bm \rho\, (k=1)}^{-1},$ estimated with the $k=1$ model, the entry for $(NO_2, PM_{10})$ is close to 0, which might suggest that $PM_{10}$ is conditionally independent to $NO_2$ given $NO.$ From the fitted model with $k=2$ components, we have $\hat{\bm \mu}_2 = (0.83, 0.90, 2.73),$ meaning that the second mixture component is further in the tail region, as all $\mu_2^i>0$ for $i=1,2,3.$ In addition, the entry for $(NO_2, PM_{10})$ of $\Sigma_{\bm \rho, 1\, (k=2)}^{-1}$ remains close to 0, suggesting that $PM_{10}$ might be conditionally independent of $NO_2$ given $NO$ in the body of the data. However, in $\Sigma_{\bm \rho, 2\, (k=2)}^{-1},$ the entry for $(NO_2, PM_{10})$ is no longer close 0, whereas the entry for $(NO, PM_{10})$ is. This might indicate that variable $PM_{10}$ is potentially conditionally independent to $NO$ given $NO_2$ in the extremes. This interpretation would closely agree with the pairwise analysis given that $(NO, PM_{10})$ are potentially AI. This conclusion would benefit from input from atmospheric scientists as it would be reassuring to know if there was a physical basis for the transition of conditional independence from the body to the tails of the joint distribution.

\begin{table}[h!]
    \centering
    \caption{Off diagonal values of the estimated precision matrices $\Sigma^{-1}_{\bm \rho\, (k=1)}$ and $\Sigma^{-1}_{\bm \rho, j\, (k=2)}$ for $j=1,2$ for triplet $(NO_2, NO, PM_{10}).$ The values considered close to 0 are highlighted in bold. All values are rounded to 2 decimal places.}
    \begin{tabular}{cc|ccc}
        \multicolumn{2}{c|}{Model} & $\Sigma^{-1}_{(NO_2, NO)}$ & $\Sigma^{-1}_{(NO_2, PM_{10})}$ & $\Sigma^{-1}_{(NO, PM_{10})}$\\
        \hline
        \multicolumn{2}{c|}{$k = 1$} & $-1.80$ & $\bm{-0.09}$ & $-0.92$ \\
        \hline
        \multirow{2}{*}{$k = 2$} & $(j = 1)$ & $-1.61$ & $\bm{-0.07}$ & $-0.43$ \\
        & $(j = 2)$ & $-1.68$ & $-0.73$ & $\bm{0.03}$ \\
        \hline
    \end{tabular}
    \label{tab:precd3}
\end{table}

\subsection{Higher dimensional analysis}  \label{subsection:fulld5}

\begin{figure}[h!]
    \centering
    \includegraphics[width=0.9\textwidth]{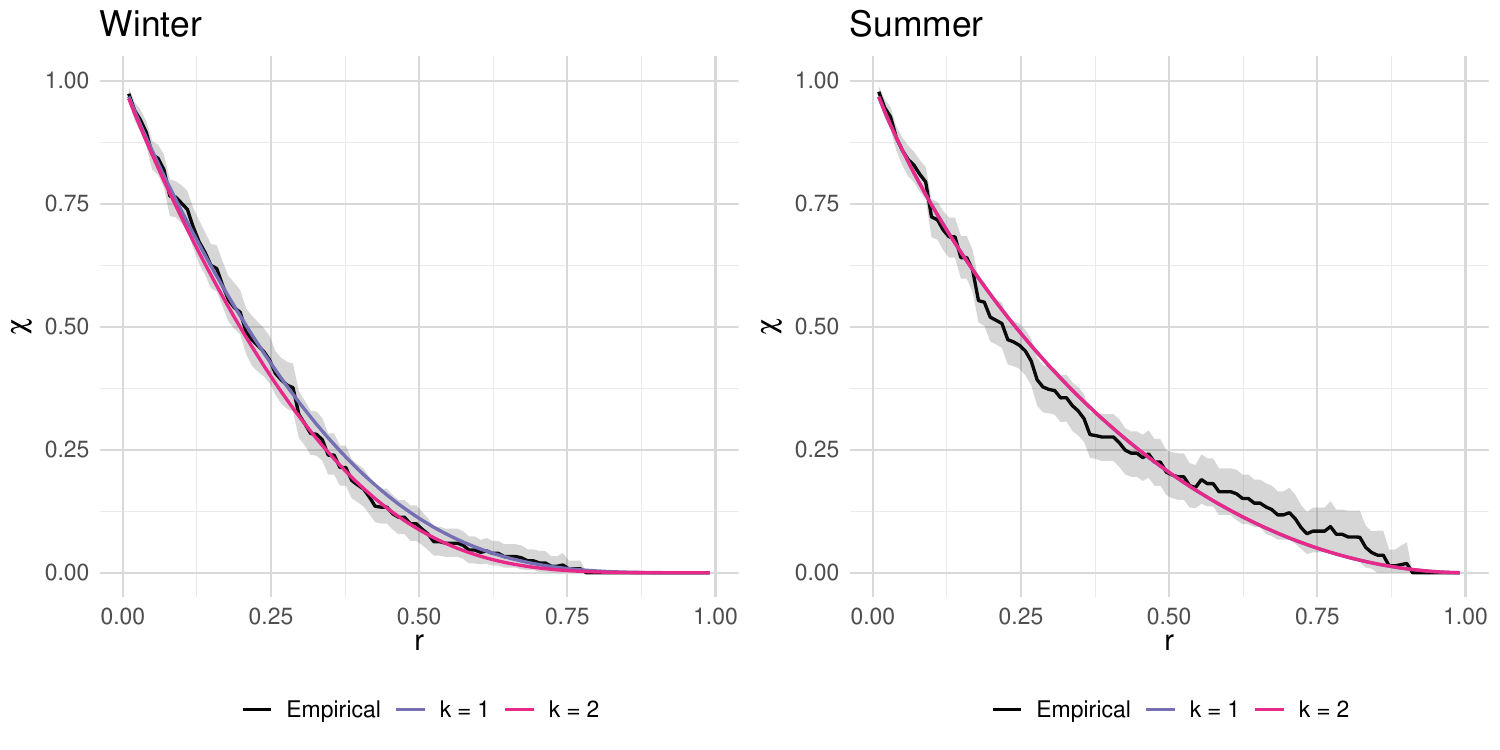}
    \caption{Estimates of $\chi_5(r)$ for $r\in (0.1)$ with empirical (in black) values also shown for $(O_3, NO_2, NO, SO_2, PM_{10})$ in the winter season (left) and the summer season (right). The pointwise $95\%$ confidence intervals for the empirical $\chi_5(r)$ are obtained through bootstrap. Note that $\chi_5(r)$ for $k=1$ and $k=2$ overlap in the right panel.}
    \label{fig:appd5}
\end{figure}

A similar analysis is performed for the full data set $(d = 5),$ where, contrary to the pairwise and trivariate analysis, the summer season is also studied. In this case, when considering all the pollutants jointly, the Gaussian mixture copula with only $k=1$ is the preferred one. In particular, for the summer season, a mixing probability $\hat p_1$ of exactly one is obtained when considering $k=2$ components, meaning that adding an extra component only adds complexity to the model. This is visible in Table~\ref{tab:appd5} and Figure~\ref{fig:appd5} with the changes in AIC values and model-based $\chi_5(r)$ obtained. For the summer season, the $k=2$ model reduces to the $k=1$ model according to the estimated mixing probabilities, with the larger number of parameters reflected on the change in AIC. 

\begin{table}[h!]
    \centering
    \caption{Change in AIC values obtained for the Gaussian mixture copula for $k=2$ relative to when $k=1$ for $(O_3, NO_2, NO, SO_2, PM_{10})$ for the winter and summer seasons. The mixing probabilities $(\hat p_1, \hat p_2)$ are reported for the $k=2$ model. All the values are rounded to 3 decimal places.}
    \begin{tabular}{l|c|c}
         Season &  $\text{AIC}_{k_1-k_2}$ & $(\hat p_1, \hat p_2)$\\
         \hline
        Winter & $25.519$ & $(0.997, \, 0.003)$  \\
        Summer & $40.773$ & $(1.000, \, 0.000)$  \\
        \hline
    \end{tabular}
    \label{tab:appd5}
\end{table}

Exploring the graphical structure of the underlying data could help reducing the dimensionality in such cases. In particular, potentially conditional independence between variables could be taken into account during the analysis. We report the off-diagonal values of the estimated precision matrices from the $k=1$ and $k=2$ models for the winter and for $k=1$ in summer in Table~\ref{tab:precd5}. For the winter, the entry for $(NO, SO_2)$ of $\Sigma_{\bm \rho\, (k=1)}^{-1}$ is close to 0, which might suggest that $NO$ and $SO_2$ are conditionally independent given $O_3,$ $NO_2$ and $PM_{10}.$ The same entry remains close to 0 for the first mixture component $j=1$ from the $k=2$ model, which would still indicate that these variables are conditionally independent given the remaining pollutants. In addition, the entry $(NO, PM_{10})$ of $\Sigma_{\bm \rho, 2\, (k=2)}^{-1}$ is near 0, which suggests that, given the remaining variables, $NO$ and $PM_{10}$ are potentially conditionally independent further in the tail. However, given that $\hat{\bm \mu}_2=(2.36, 3.18, -0.20, 0.16, 3.96)$ and thus not all $\mu_2^i>0$ for $i\in D,$ the second mixture component is essentially capturing asymmetry and not a difference in body and tail dependence. For summer, the results suggest that $O_3$ and $SO_2$ may be conditionally independent given $NO_2,$ $NO$ and $PM_{10},$ and we also find that $(NO, SO_2)$ might be conditionally independent given the remaining variables across both seasons.

\begin{table}[h!]
    \renewcommand{\arraystretch}{1.15}
    \centering
    \caption{Off diagonal values of the estimated precision matrices $\Sigma^{-1}_{\bm \rho\, (k=1)}$ and $\Sigma^{-1}_{\bm \rho, j\, (k=2)}$ for $j=1,2$ for $(O_3, NO_2, NO, SO_2, PM_{10}).$ The values considered close to 0 are highlighted in bold. All values are rounded to 2 decimal places.}
    \begin{tabular}{l|cc|ccccc}
         &\multicolumn{2}{c|}{Model} & $\Sigma^{-1}_{(O_3, NO_2)}$ & $\Sigma^{-1}_{(O_3, NO)}$ & $\Sigma^{-1}_{(O_3, SO_2)}$ & $\Sigma^{-1}_{(O_3, PM_{10})}$ & $\Sigma^{-1}_{(NO_2, NO)}$ \\
        \hline
        \multirow{3}{*}{W} & \multicolumn{2}{c|}{$k = 1$} & $-0.75$ & $0.88$ & $0.68$ & $0.14$ & $-1.86$ \\
        \cline{2-8}
        & \multirow{2}{*}{$k = 2$} & $(j = 1)$ & $-0.81$ & $0.91$ & $0.66$ & $0.38$ & $-2.03$ \\
        & & $(j = 2)$ & $-0.31$ & $0.33$ & $0.11$ & $0.60$ & $0.25$ \\
        \hline
        S & \multicolumn{2}{c|}{$k = 1$} & $-0.54$ & $0.80$ & $\bm{0.00}$ & $-0.24$ & $-1.45$ \\
        \hline\hline
         &\multicolumn{2}{c|}{Model} & $\Sigma^{-1}_{(NO_2, SO_2)}$ & $\Sigma^{-1}_{(NO_2, PM_{10})}$ & $\Sigma^{-1}_{(NO, SO_2)}$ & $\Sigma^{-1}_{(NO, PM_{10})}$ & $\Sigma^{-1}_{(SO_2, PM_{10})}$ \\
        \hline
        \multirow{3}{*}{W} & \multicolumn{2}{c|}{$k = 1$} & $-0.38$ & $-0.38$ & $\bm{0.08}$ & $-0.58$ & $-0.44$ \\
        \cline{2-8}
        & \multirow{2}{*}{$k = 2$} & $(j = 1)$ & $-0.18$ & $-0.20$ & $\bm{0.02}$ & $-0.58$ & $-0.37$ \\
        & & $(j = 2)$ & $-0.20$ & $0.20$ & $0.90$ & $\bm{0.04}$ & $0.42$ \\
        \hline
        S & \multicolumn{2}{c|}{$k = 1$} & $-0.48$ & $-0.40$ & $\bm{0.03}$ & $-0.28$ & $-0.54$ \\
        \hline
    \end{tabular}
    \label{tab:precd5}
\end{table}

From Figure~\ref{fig:appd5}, it is clear that $\chi_5(r)\to 0$ as $r\to 1$ when considering the joint behaviour of all pollutants for both seasons indicating that all the pollutants cannot be large together; given that some pollutants are AI between pairs, this is not surprising. The model-based $\chi_5(r)$ obtained with both $k=1$ and $k=2$ lie within pointwise $95\%$ confidence intervals for the empirical estimate of $\chi_5(r),$ especially in the winter season. Moreover, both model $\chi_5(r)$ estimates are close to the empirical values, indicating that either model is a good fit to the data. Although not as pronounced as in the winter season, similar conclusions can be drawn for the summer season. The corresponding results for $\eta_5(r)$ are presented in Figure~\ref{fig:appd5sup} of the Supplementary Material. Whilst for the summer season, the model estimates of $\eta_5(r)$ approach 0.35 as $r\to 1,$ for the winter season $\eta_5(r)\to 0.2$ with the $k=1$ model, and $\eta_5(r)\to 0.15$ with the $k=2$ model. These results indicate that, in the summer season, the extremes of $(O_3, NO_2, NO, SO_2, PM_{10})$ are positively dependent, but in the winter season, they either nearly independent according to the $k=1$ model, or negatively dependent based on the mixture model with $k=2$ components. We note that there are no points that are jointly bigger than $r>0.75,$ which results in $\eta_5(r)$ not being defined (recall expression~\eqref{eq:etarddim}). Thus, a drop in the empirical $\eta_5(r)$ and corresponding pointwise confidence intervals values is observed.

\begin{figure}[h!]
    \centering
    \includegraphics[width=0.9\textwidth]{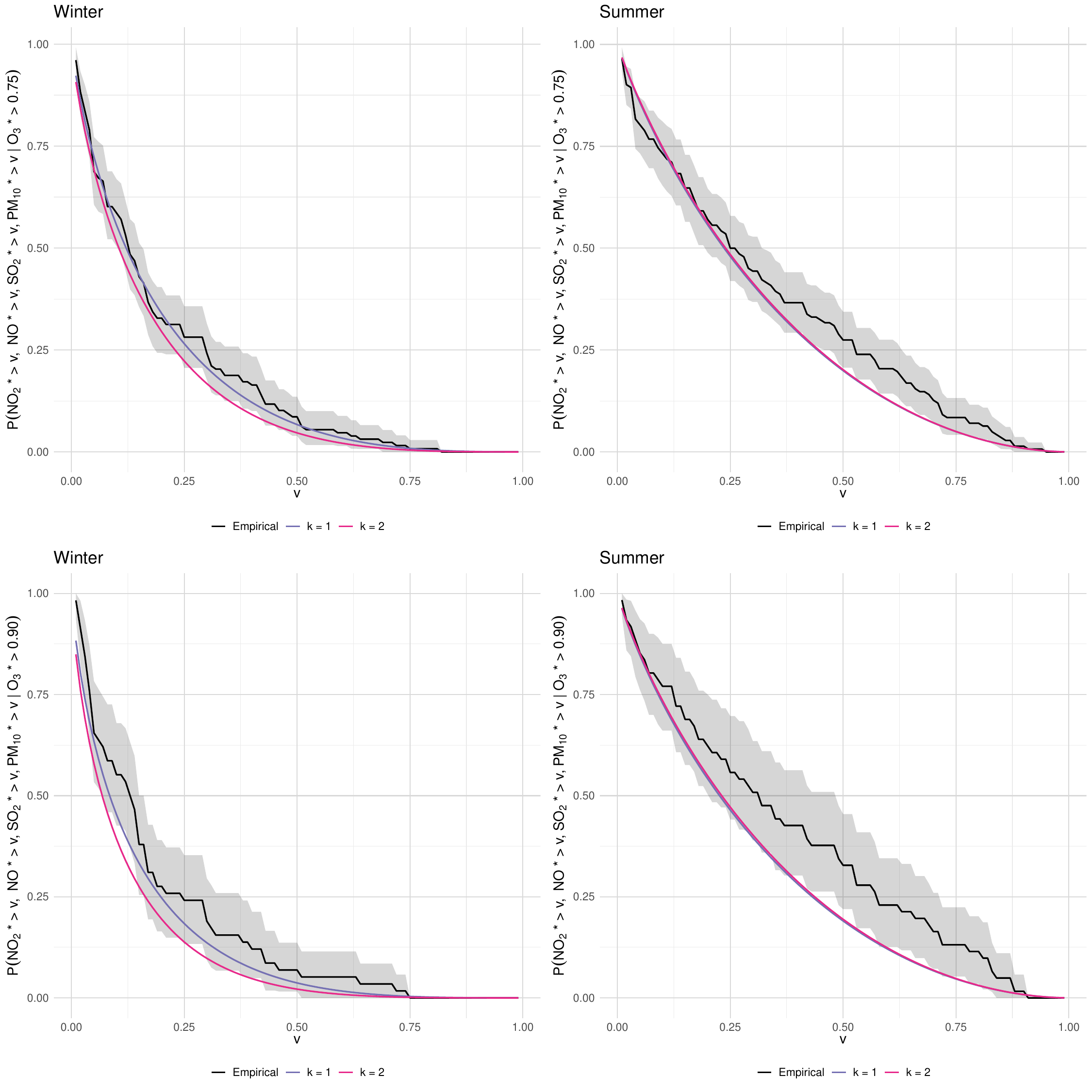}
    \caption{Comparison between model-based probabilities $\Pr(NO^*_2 >v, NO^* > v, SO^*_2 >v, PM^*_{10} > v \mid O^*_3 > u)$ for $v\in (0,1),$ for two large values $u=\{0.75, 0.90\}$ given by the Gaussian mixture copula with $k=1$ (in purple) and $k=2$ (in pink) components. The empirical probability is given in black, and its pointwise 95$\%$ confidence intervals are obtained through bootstrap. Note that for the summer season, the $k=1$ and $k=2$ model probabilities overlap.}
    \label{fig:probd5}
\end{figure}

We assess the performance of the Gaussian mixture copula by considering the behaviour of the remaining variables when conditioning for $O_3^*$ being larger than $u=\{0.75, 0.90\},$ and compare the model-based probabilities that the other variables are each larger than $v\in (0,1)$ with their empirical counterpart; these are shown in Figure~\ref{fig:probd5}. For each season, the fitted models seem to capture the conditioning behaviour for all levels $v,$ especially when $u=0.75$. However, for $u=0.9,$ and particularly for the summer season, there is evidence that the model fit can be improved as the probabilities estimated by the model lie outside the pointwise $95\%$ confidence intervals.

\section{Conclusions and discussion} \label{section:conclusion}

We proposed a copula model based on a mixture of multivariate Gaussian distributions to represent the body and tail regions of multivariate data. This copula model avoids the need to specify a threshold vector which defines an extremal region, and is able to represent a broad range of complex extremal dependence structures. Theory and simulation studies showed that the Gaussian mixture copula is able to capture asymptotic dependence at arbitrary large quantiles, particularly for models with $k = 3$ mixture components, or $k =2$ with larger sample sizes. Additionally, we showed that the Gaussian mixture copula is flexible enough to fit more complex data structures, including non-exchangeable data; in particular, the model captures the sub-asymptotic joint tail behaviour along different rays accurately. 

We showcased the performance of the Gaussian mixture model by applying it to the $5$-dimensional seasonal air pollution data set analysed by \citet{HeffernanTawn2004}. We started by performing a bivariate analysis on the pairs of pollutants identified by \citet{HeffernanTawn2004} as exhibiting asymptotic dependence. When applying the proposed copula model, we obtained similar findings for sub-asymptotic levels; more specifically, we showed that using a model with $k=2$ mixture components, the joint behaviour could be effectively characterised well into the tails. In higher dimensions, it becomes evident that the fitted Gaussian mixture copula exhibits asymptotic independence, which is consistent with the empirical evidence based on non-parametric estimates. Nevertheless, it provides a more accurate representation of the joint behaviour with $k=2$ components when compared to $k=1.$ This conclusion was further supported by examining the conditional behaviour of the variables at various levels, given one variable being large. For each analysis, we constructed the copula model based on the dimension of each data set. Alternatively, it would be interesting to evaluate the performance of the copula model in fitting the pairs and triple by marginalising the 5-dimensional copula over the variables not included in the joint vector of interest. Finally, as shown by the pairwise study and noted by \citet{Simpsonetal2020}, there are subsets of variables that exhibit asymptotic dependence, even though the full joint vector exhibits asymptotic independence.

Although the Gaussian mixture copula scales relatively well to higher dimensions, the evaluation of its log-likelihood becomes increasingly computationally expensive when $d\geq 2$. For instance, when moving from a bivariate to a $5$-dimensional setting, our simulation studies showed that the computational time increased noticeably for a model with $k=2$ mixture components. This is heavily due to the high number of correlation parameters in the model, but also due to the need for inversion of functions when constructing the copula model. These issues lead to complications in the inference procedure, particularly when we wish to consider adding an extra mixture component, or moving to an even higher dimensional setting. Since simulation from the model is straightforward and efficient, the computational burden of the inference procedure can be mitigated by employing simulation-based methods. Such methods, often referred to as likelihood-free approaches, do not rely on the knowledge of a likelihood function. Examples include approximate Bayesian computation (ABC; e.g., \citealp{Sissonetal2018}) or neural network-based techniques (e.g., \citealp{Zammitetal2024}). Alternatively, the number of parameters in the model could be reduced by exploring data reduction methods for the covariance structure, such as those used in the Gaussian mixture models considered by \citet{McNicholasMurphy2008}.

\bigskip

\noindent {\bf Declarations of Interest:} None.

\section*{Acknowledgments}
This paper is based on work completed while L\'idia Andr\'e was part of the EPSRC funded STOR-i Centre for Doctoral Training (EP/S022252/1).

\bibliography{ref}{}
\bibliographystyle{apalike}


\title{Supplementary Material for \emph{Gaussian mixture copulas for flexible dependence modelling in the body and tails of joint distributions}}
\author{L. M. Andr\'e$^{1*}$ and J. A. Tawn$^{2}$\\
\small $^{1}$ Namur Institute for Complex Systems, University of Namur, Rue Graf\'e 2, Namur 5000, Belgium\\
\small $^{2}$ School of Mathematical Sciences, Lancaster University, LA1 4YF, United Kingdom \\
\small $^*$ Correspondence to: \href{mailto:lidia.andre@unamur.be}{lidia.andre@unamur.be} }
\date{}

\maketitle
\pagenumbering{arabic}

\setcounter{section}{0}
\setcounter{figure}{0}
\setcounter{table}{0}

\renewcommand{\thefigure}{S\arabic{figure}}
\renewcommand{\thetable}{S\arabic{table}}
\renewcommand{\theequation}{S\arabic{equation}}
\renewcommand{\thesection}{S\arabic{section}}

\section{Formulation of set $A_w$ from Section~\ref{subsubsection:overview}}\label{supsec:appendixpaper2}

Here we present only the bivariate case, with the general $d$-dimensional case following similarly. Consider $w\in\mathcal{S}_1,$ and standard exponential random variables, $X_1^E$ and $X_2^E,$ with marginal distribution function $F_{E}(x)=1-\exp\{-x\},$ for $x>0$ and $i=1, 2.$ Following \citet{WadsworthTawn2013}, we are interested in regions of the form 
\begin{equation}\label{eq:setaxy}
    A(x, y) =\{X_1^E > x, X_2^E > y\}, \quad \text{for }x>0 \text{ and } y>0.
\end{equation}

\begin{figure}[H]
    \centering
    \includegraphics[width=0.8\textwidth]{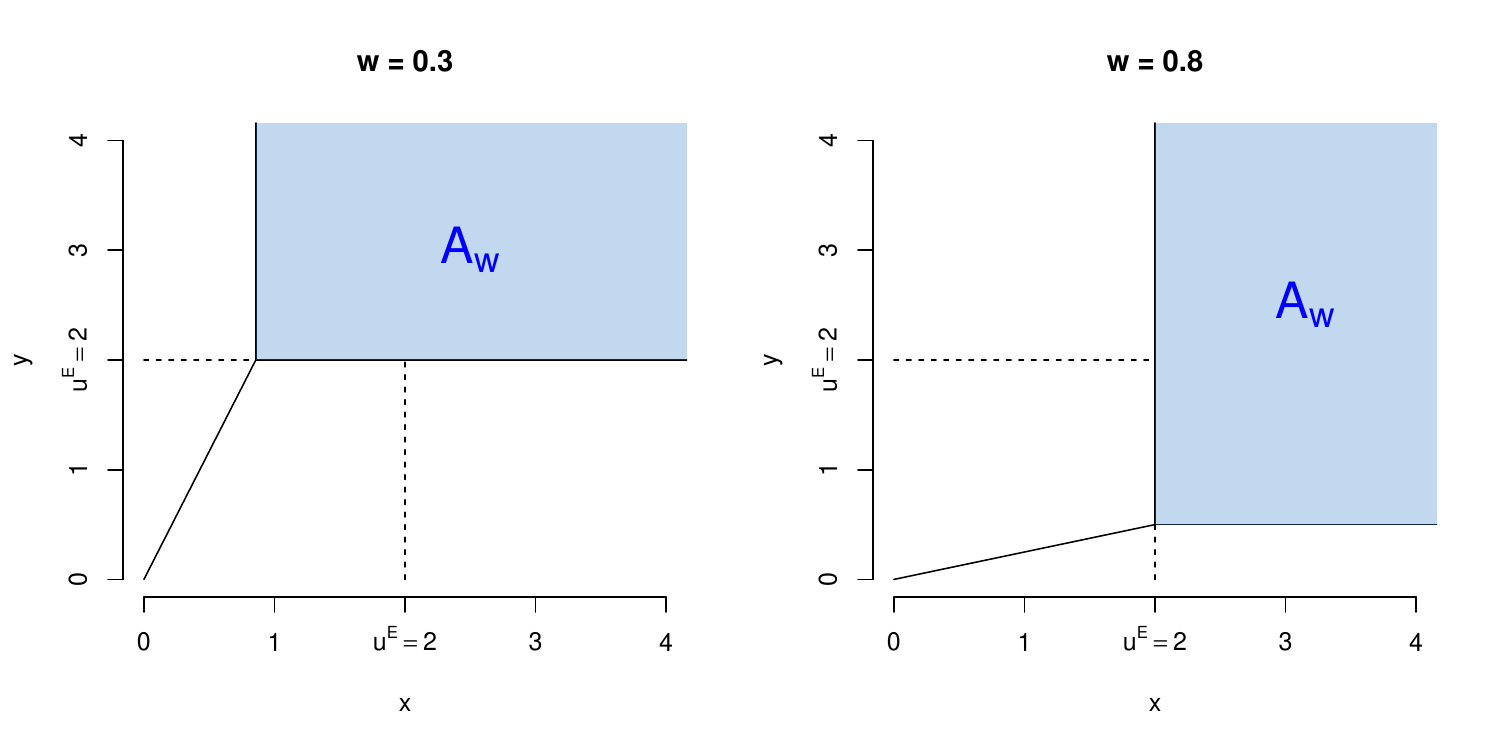}
    \caption{Example of regions $A_w$ for $w=\{0.3, 0.8\}$ and $u^E=2.$}
    \label{fig:appendixAw}
\end{figure}

Let us now assume that $w:=x\slash (x+y)$ and $\max\{x, y\} = u^E,$ where $u^E$ is some threshold level in exponential margins. Two examples of such sets are shown by the shaded regions in Figure~\ref{fig:appendixAw}.

By the definition of $w,$ we have that
\begin{equation}
    \max\{x, y\} = x \Leftrightarrow \max\left\{x, \frac{1-w}{w}x\right\} = x \Rightarrow 1 > \frac{1-w}{w} \Leftrightarrow w > \frac{1}{2}.
\end{equation}
Thus, we have $x=u^E$ and set $A(x,y)$ from expression~\eqref{eq:setaxy} can be rewritten as
\begin{equation}
    A_w=\left\{X_1^E > u^E, X_2^E > \frac{1-w}{w}u^E\right\}, \quad \text{when }w> 1\slash 2. 
\end{equation}

Similarly, we have $w\leq 1\slash 2$ when $\max\{x,y\}=y=u^E.$ Therefore,
\begin{equation}
    A_w=\left\{X_1^E > \frac{w}{1-w}u^E, X_2^E > u^E\right\},  \quad \text{for }w\leq 1\slash 2.
\end{equation}
Combining the two, we arrive to region given in Section~\ref{subsubsection:overview}.
\begin{equation}
    A_w=\left\{X_1^E > \max\left\{\frac{w}{1-w}, 1\right\}u^E, X_2^E > \max\left\{\frac{1-w}{w}, 1\right\}u^E\right\}.
\end{equation}

\newpage

\section{Implementation details} \label{supsec:modeldef}


The identifiability constraints on the parameters are imposed within the log-likeli\-hood function~\eqref{eq:loglike} of the main paper. More specifically, every time the optimisation algorithm evaluates a parameter value that fails to satisfy the constraints, a value of $\ell(\bm \theta)$ of $-\infty$ is returned. In the case of the mixing probabilities, the estimated $p_{k}$ $(k \geq 2)$ is obtained implicitly as $\bm p\in \mathcal{S}_{d-1}.$ In a higher-dimensional setting $(d\geq 2),$ the optimisation of the log-likelihood~\eqref{eq:loglike} is initially performed in a lower-dimensional setting to ensure (faster) convergence to a global maximum; specifically, all pairwise parameter estimates are obtained, and then these are used as initial values for the parameters in the higher-dimensional optimisation. When these initial values do not meet higher dimensional constraints, for example leading to non semi-positive definite $\Sigma_j,$ small perturbations on $\bm \rho_{\Sigma_j}$ $(j\in K)$ are added.

\section{Simulation Studies} \label{supsec:simulation}

\subsection{Model inference} \label{supsubsec:modelinf}

We showcase the identifiability and inference performance of the Gaussian mixture copula by performing a simulation study, illustrating the performance of the sampling distribution of the MLE of $\bm \theta$ over i.i.d. \!replicated samples. To do so, we consider three Gaussian copula model specifications with $(d,k)=(2,2)$ (Case I), $(d,k)=(2,3)$ (Case II) and $(d,k)=(5, 2)$ (Case III) with parameters denoted by $\bm \theta_{\text{I}},\, \bm \theta_{\text{II}}$ and $\bm \theta_{\text{III}},$ respectively. In all cases, i.i.d. \!realisations from model~\eqref{eq:copdens} from the main paper are generated with a sample size of 1000, and each sample is simulated 50 times. Examples of Cases I-III with pairwise exchangeability, given in Figure~\ref{fig:supiden}, indicate that identifiability is not an issue when a simplified model specification is assumed with most estimates concentrated around the truth for all cases. 

\begin{figure}[h!]
    \centering
    \begin{subfigure}[b]{0.49\textwidth}
        \includegraphics[width=\textwidth]{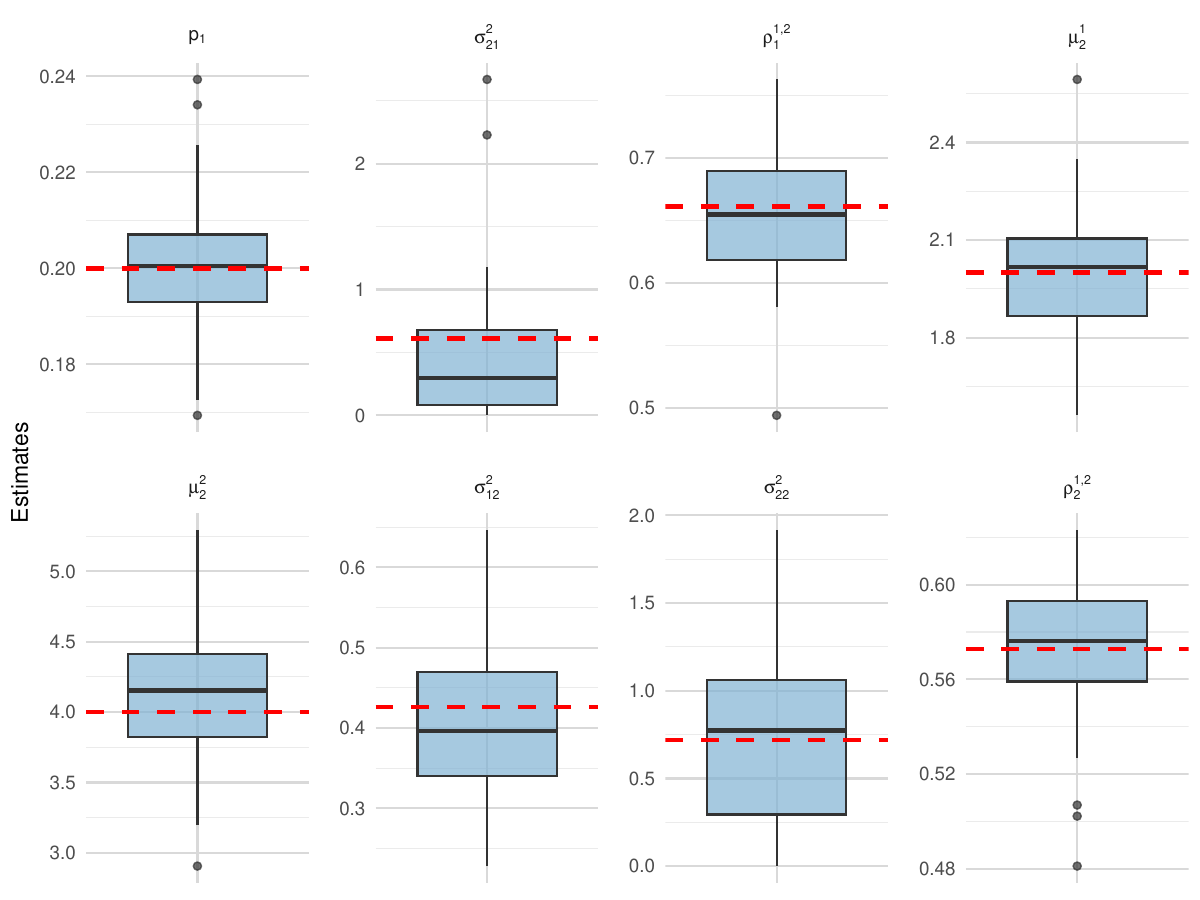}
        \caption{}
        \label{subfig:d2k2}
    \end{subfigure}
    \hfill
    \begin{subfigure}[b]{0.49\textwidth}
        \includegraphics[width=\textwidth]{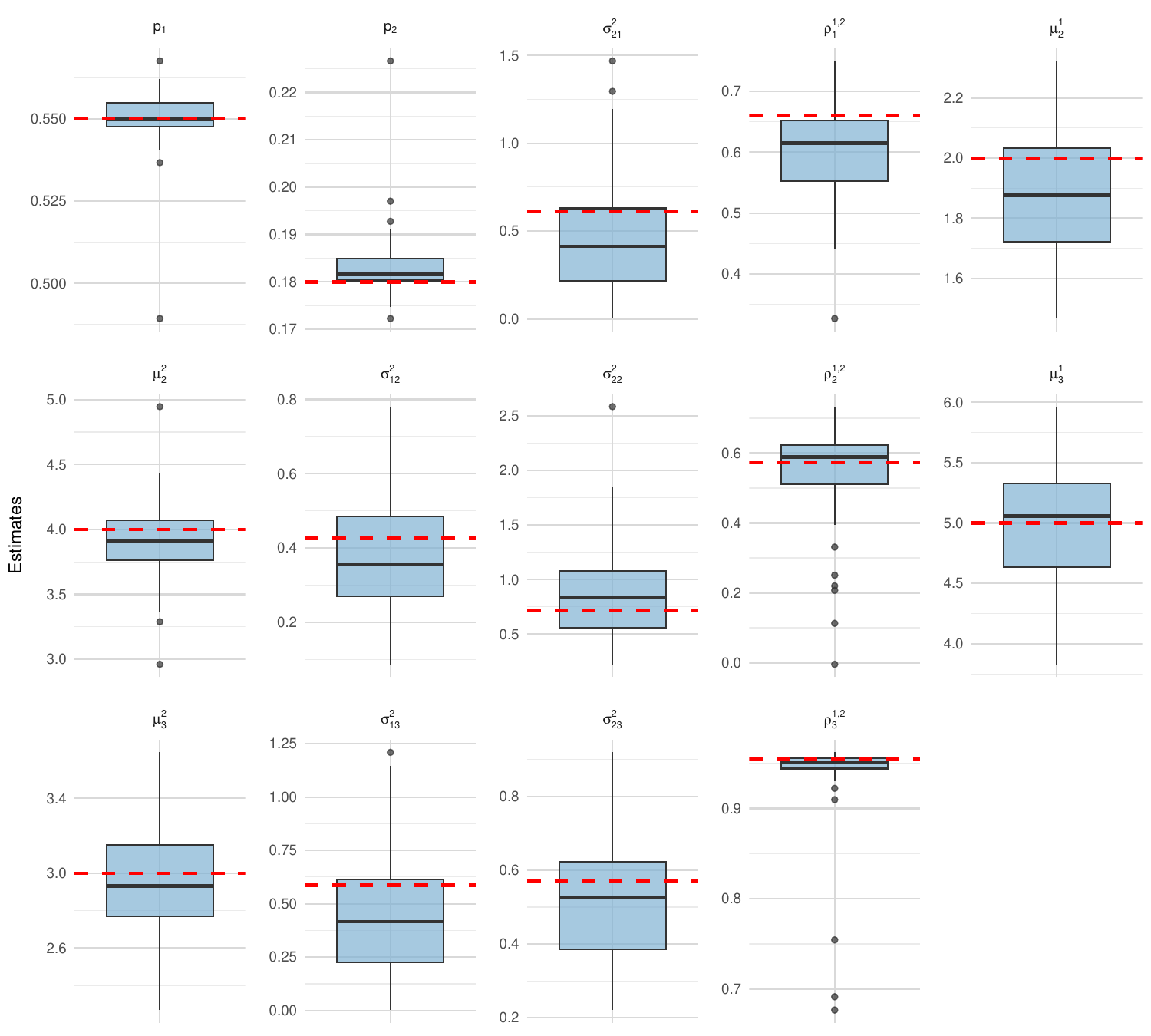}
        \caption{}
        \label{subfig:d2k3}
    \end{subfigure}
    \caption{Boxplots of estimates of the Gaussian mixture copula model based on 50 replicated data sets: (a) Case I and (b) Case II. The true parameter values are indicated by the red lines.}
    \label{fig:cased2}
\end{figure}

\begin{figure}[h!]
    \centering
    \includegraphics[width=\textwidth]{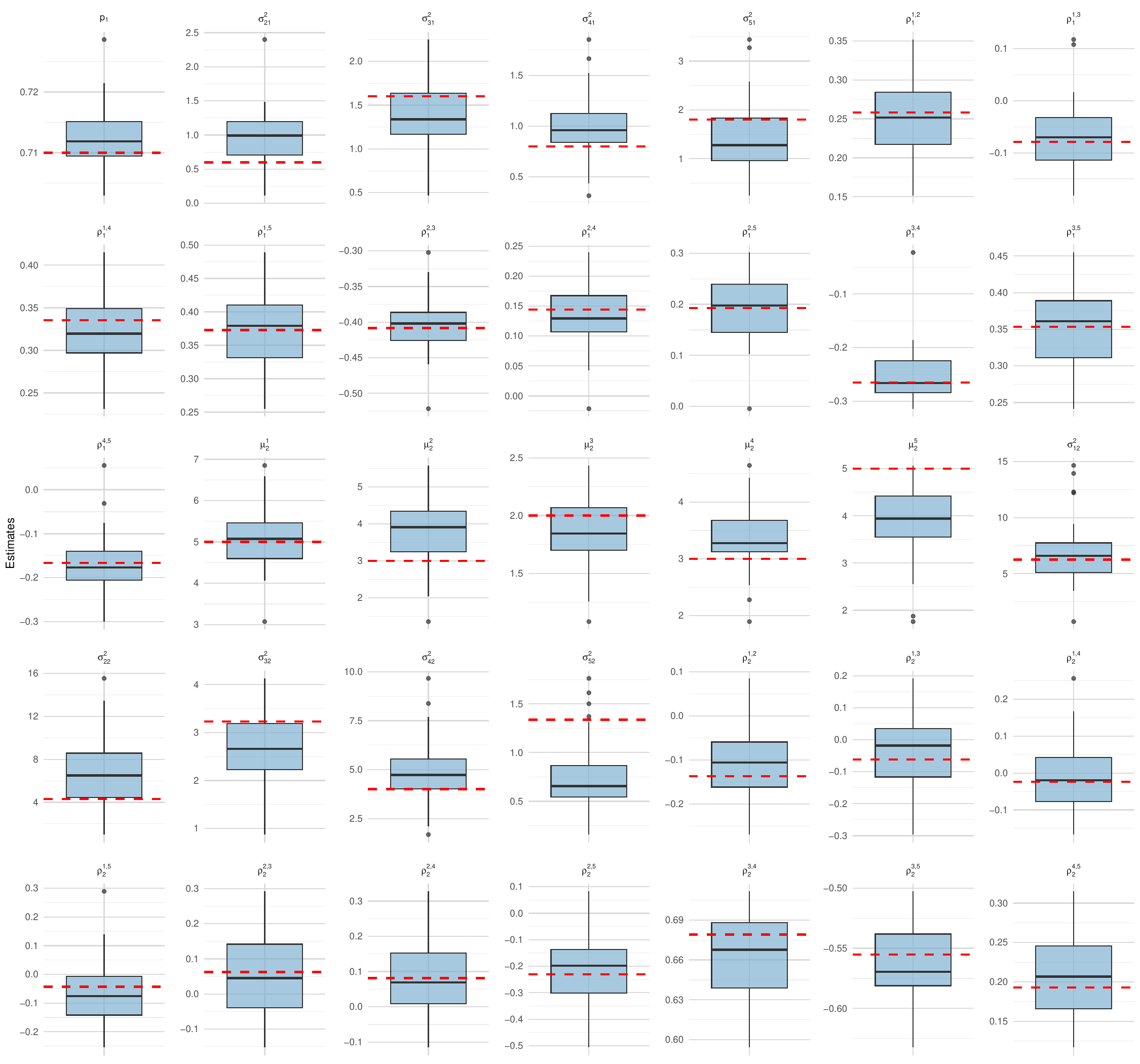}
    \caption{Boxplots of estimates of the Gaussian mixture copula model based on 50 replicated data sets for Case III. The true parameter values are indicated by the red lines.}
    \label{fig:cased5sup}
\end{figure}

For Case I, we set $p_1 = 0.20,$ $\bm \mu_1=\bm 0,$ $\bm \mu_2 = (2, 4),$ $\bm \sigma_{\Sigma_1} = (1.00, 0.61),$ $\bm \sigma_{\Sigma_2} = (0.43, 0.72),$ $\rho_{\Sigma_1} = 0.66$ and $\rho_{\Sigma_2} = 0.57.$ In Case II, when an extra mixture component is added, we retain the models for the $\bm Z_1$ and $\bm Z_2$ mixture components, and for the extra mixture component we take $(p_1, p_2) = (0.55, 0.18),$ $\bm \mu_3 = (5,3),$ $\sigma_{\Sigma_3}=(0.59, 0.57)$ and $\rho_{\Sigma_3} = 0.96.$ For Case III, we take $p = 0.71,$ $\bm \mu_1 = \bm 0,$ $\bm \mu_2 = (5,3,2,3,5),$ $\bm \sigma_{\Sigma_1} = (1.00, 0.60, 1.60, \linebreak 0.80, 1.80),$ $\bm \sigma_{\Sigma_2} = (6.26, 4.31, 3.23, 4.01, 1.34),$ $\bm \rho_{\Sigma_1} = (0.26, -0.08, 0.34, 0.37, -0.41, 0.14, \linebreak 0.19, -0.27, 0.35, -0.17)$ and $\bm \rho_{\Sigma_2} = (-0.14, -0.06, -0.02, -0.04, 0.06, 0.08, -0.23, 0.68, \linebreak -0.56, 0.19).$ Figure~\ref{fig:cased2} shows the results of the simulation study for Cases I and II in the left and right panels, respectively, and the results for Case III are displayed in Figure~\ref{fig:cased5sup}. 

From the findings of Cases I-III, there is no indication that model identifiability is a concern. It can be seen that the MLE estimates seem to be concentrated around the true values for most parameters, particularly when the model includes fewer parameters. This is to be expected since less parameters often leads to smaller variability in the estimation and parameter dependencies. When moving to a higher dimensional setting, estimation becomes computationally more expensive. Furthermore, as shown in Figure~\ref{fig:cased5sup}, a few of the estimates appear to deviate further from the true values, particularly those associated with the $\bm Z_2$ mixture component, which seem to show higher variability. Given the high number of parameters to estimate, and that the numerical maximiser converged, without any convergence concerns, for all the 35 parameters in the model, this is not considered an issue with the model or its parameterisation.

\begin{figure}[h!]
    \centering
    \includegraphics[width=\textwidth]{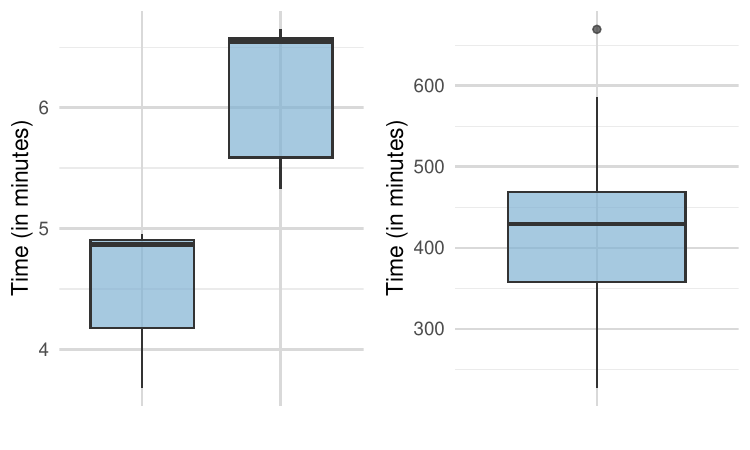}
    \caption{Time (in minutes) taken to optimise the log-likelihood~\ref{eq:loglike} from the main paper of a model with $d=2$ and $k=2$ or $d=2$ and $k=3$ (left), and $d=5$ and $k=2$ (right).}
    \label{fig:times}
\end{figure}

\newpage
To assess the computational effort required to evaluate the log-likelihood function given in expression~\eqref{eq:loglike} from the main paper, especially when moving to a higher dimensional setting, we record the times taken to optimise the log-likelihood function across the three cases; these are shown in Figure~\ref{fig:times}. The log-likelihood function is evaluated using an internal computing node running Linux on an Intel Ice Lake CPU with 200GB RAM memory; see \url{https://lancaster-hec.readthedocs.io/en/latest/} for details (last accessed on 08/03/2025). As should be expected, the time to optimise one log-likelihood increases with both $d$ and $k$. While the optimisation time increases, on average, in $1.6$ minutes when one extra mixture component is added when $d=2$, in the case of a higher dimension such as $d=5$, this computational time increases in 6.9 hours, on average.

\clearpage

\subsubsection*{Pairwise exchangeability}

\begin{figure}[h!]
    \centering
    \begin{subfigure}[b]{0.49\textwidth}
        \includegraphics[width=\textwidth]{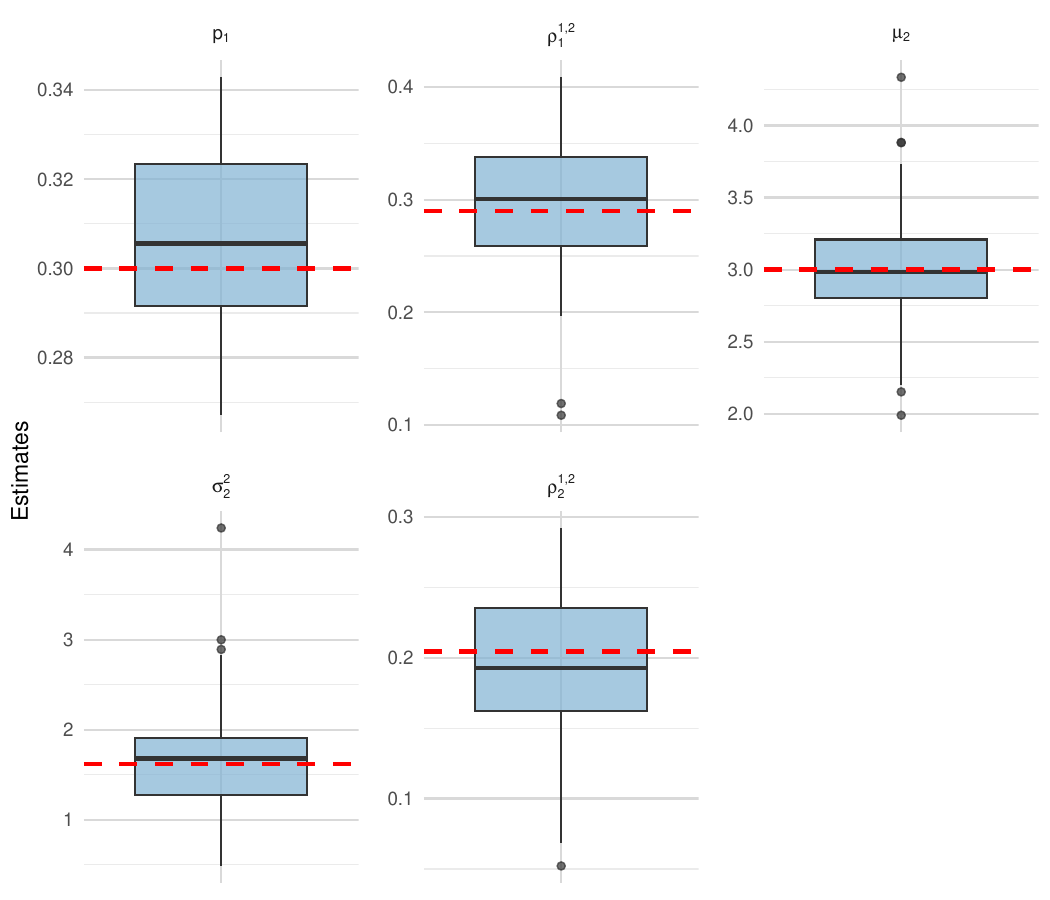}
        \caption{}
    \end{subfigure}
    \hfill
    \begin{subfigure}[b]{0.49\textwidth}
        \includegraphics[width=\textwidth]{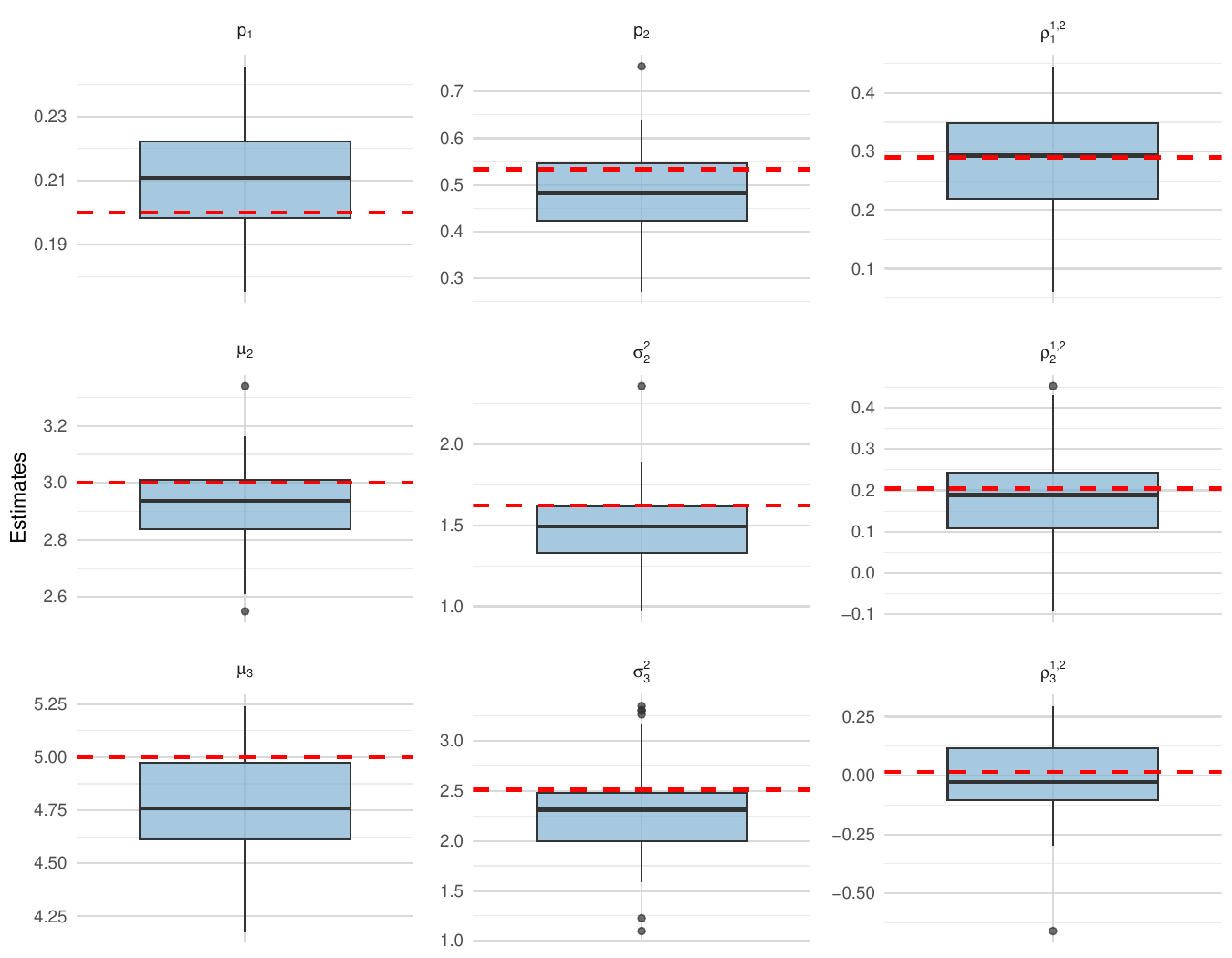}
        \caption{}
    \end{subfigure}
    \caption{Boxplots of estimates of the Gaussian mixture copula model when assuming pairwise exchangeability based on 50 replicated data sets: (a) Case I and (b) Case II. The true parameter values are indicated by the red lines.}
    \label{fig:supiden}
\end{figure}

\begin{figure}[h!]
    \centering
    \includegraphics[width=\textwidth]{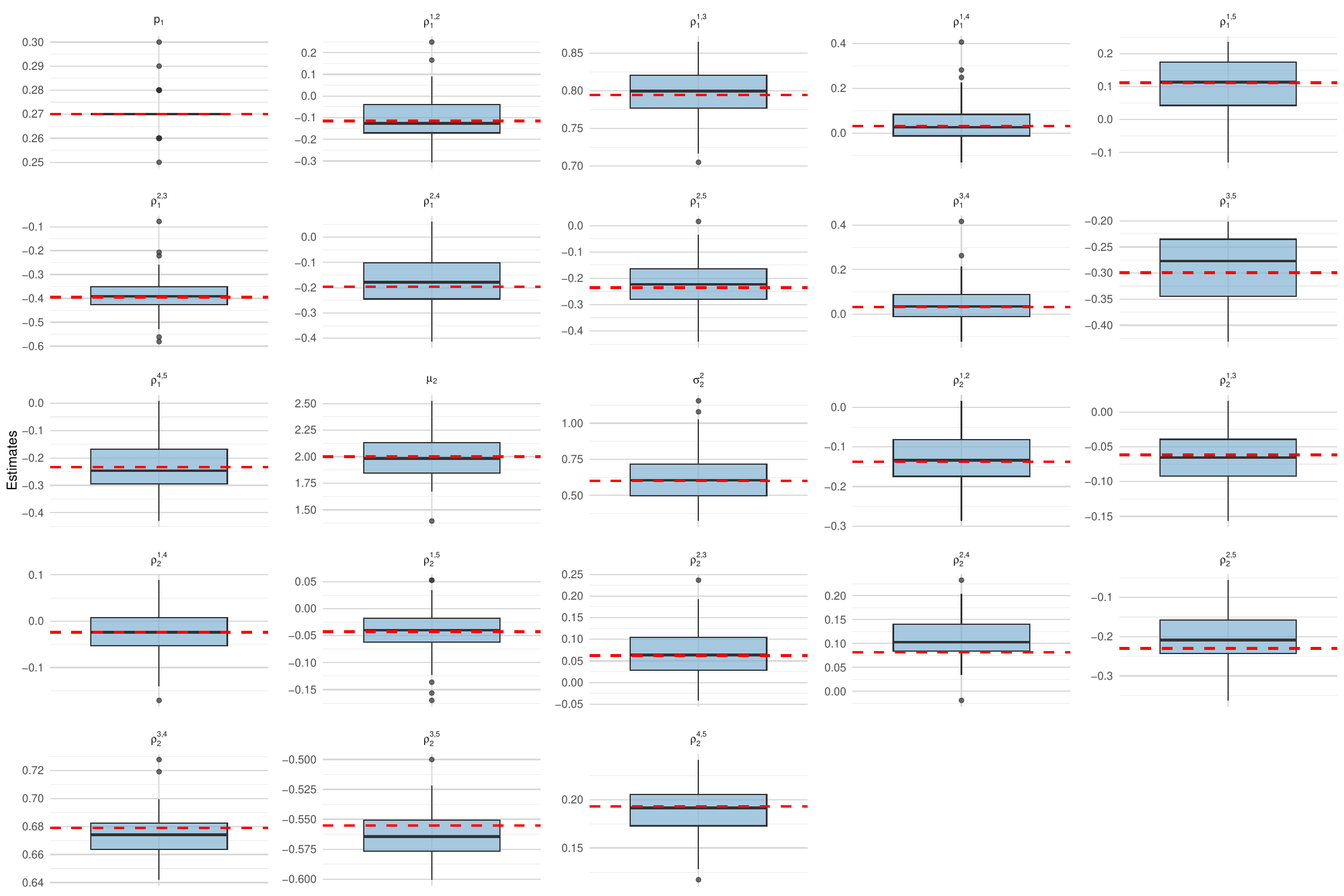}
    \caption{Boxplots of estimates of the Gaussian mixture copula model when assuming pairwise exchangeability based on 50 replicated data sets for Case III. The true parameter values are indicated by the red lines.}
    \label{fig:supidend5}
\end{figure}

We consider now pairwise exchangeability where for each mixture component $\bm Z_j,$ \linebreak $\mu_j^1=\ldots=\mu_j^d$ and $\sigma_{1j}^2=\ldots=\sigma_{dj}^2$ for $j \in K,$ $d\in D.$ For Case I, we set $p_1 = 0.30,$ $\bm \mu_1=\bm 0,$ $\bm \mu_2 = \bm 3,$ $\bm \sigma_{\Sigma_1} = \bm 1,$ $\bm \sigma_{\Sigma_2} = \bm 1.62,$ $\rho_{\Sigma_1} = 0.29$ and $\rho_{\Sigma_2} = 0.20.$ In Case II, when an extra mixture component is added, we retain the models for the $\bm Z_1$ and $\bm Z_2$ mixture components, and for the extra mixture component we take $(p_1, p_2) = (0.20, 0.53),$ $\bm \mu_3 = \bm 5,$ $\sigma_{\Sigma_3}= \bm 2.51$ and $\rho_{\Sigma_3} = 0.02.$ For Case III, we take $p = 0.27,$ $\bm \mu_1 = \bm 0,$ $\bm \mu_2 = \bm 2,$ $\bm \sigma_{\Sigma_1} = \bm 1,$ $\bm \sigma_{\Sigma_2} = \bm 0.6,$ $\bm \rho_{\Sigma_1} = (-0.12, 0.79, 0.03, 0.11, -0.39, -0.20, -0.24, 0.03, -0.30, -0.23)$ and $\bm \rho_{\Sigma_2} = (-0.14, -0.06, -0.02,  -0.04, 0.06, 0.08, -0.23, 0.68, -0.56, 0.19).$ The results are shown in Figures~\ref{fig:supiden} and \ref{fig:supidend5}, respectively, for Cases I-II and III.  

\clearpage

\subsection{Model fit and diagnostics} \label{supsubsec:modelfit}

\subsubsection{Asymptotically dependent data} \label{supsubsec:addata}

\begin{figure}[h!]
    \centering
    \includegraphics[width=0.9\textwidth]{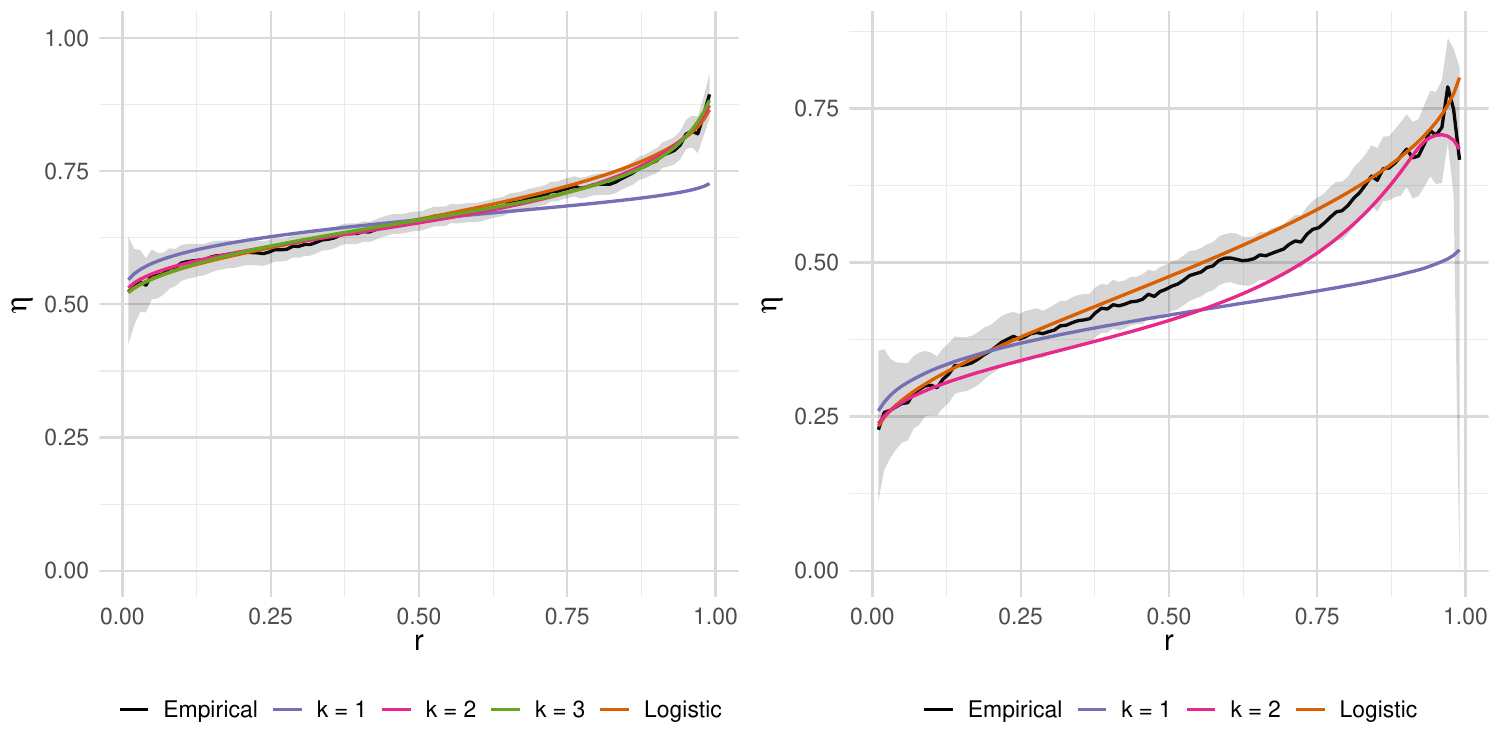}
    \caption{Estimates of $\eta_D(r)$ for $r\in (0.1)$ with true (in orange) and empirical (in black) values also shown. The pointwise $95\%$ confidence intervals for the empirical $\eta_D(r)$ are obtained through bootstrap. When $d=2$ (left), models with $k=1-3$ mixture components are considered, whereas when $d=5$ (right) only models with $k=1-2$ mixture components are studied.}
    \label{fig:adeta}
\end{figure}

Figure~\ref{fig:adeta} shows the results for $\eta_D(r)$ for the case where the underlying data is AD given in the main paper.

\begin{figure}[H]
    \centering
    \includegraphics[width = 0.9\textwidth]{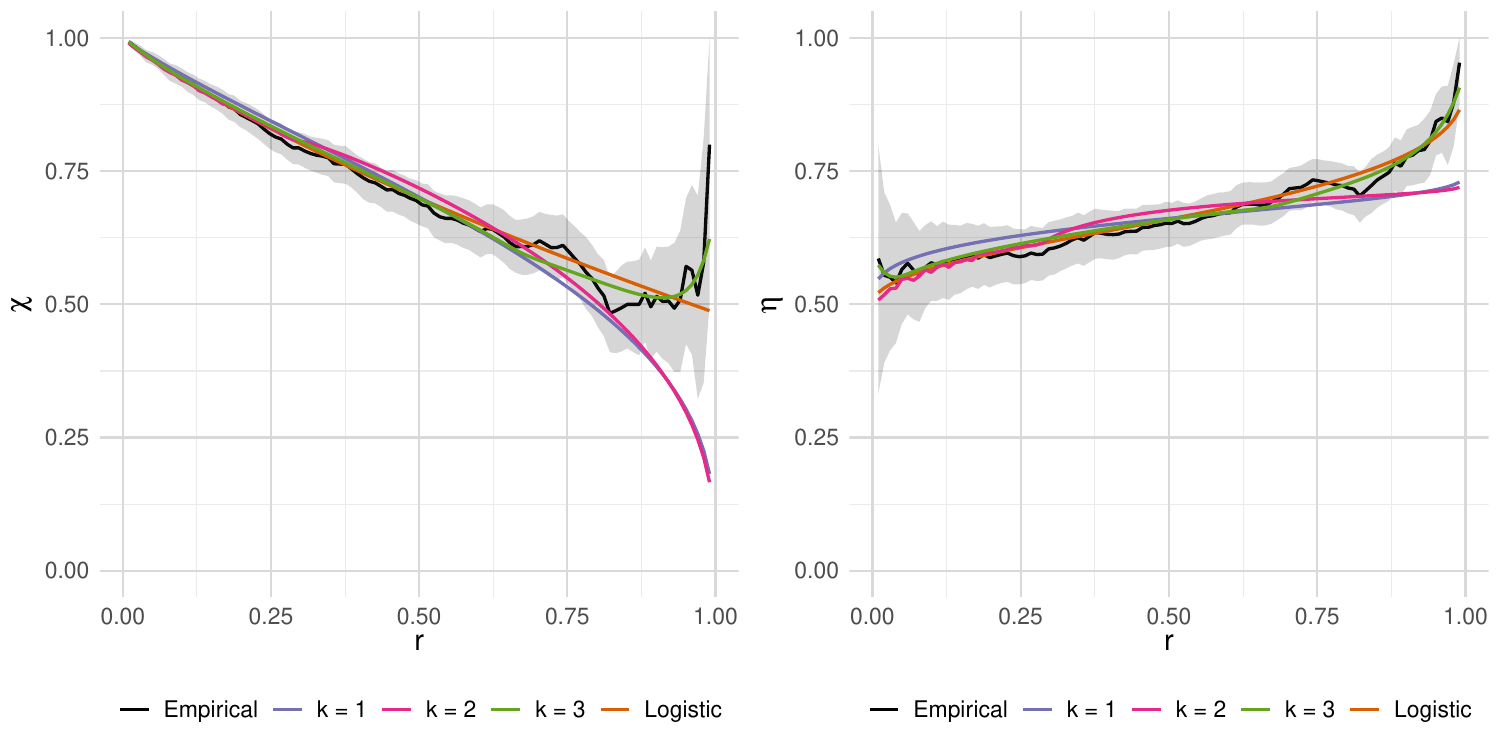}
    \caption{Estimates of $\chi_2(r)$ (left) and of $\eta_2(r)$ (right) for $r\in (0.1)$ with true (in orange) and empirical (in black) values also shown for $n=1000$. The pointwise $95\%$ confidence intervals for the empirical $\chi_2(r)$ and $\eta_2(r)$ are obtained through bootstrap.}
    \label{fig:adsup}
\end{figure}

When considering a smaller sample size $(n=1000),$ the decrease in AIC with $k=3$ in relation to when $k=1$ is $-43.18,$ whereas there is an increase in AIC of $11.69$ with $k=2$ relative to $k=1.$ These results indicate that the $k=3$ model is the most suitable for the underlying data. Figure~\ref{fig:adsup} shows a comparison between model-based $\chi_2(r)$ and $\eta_2(r)$ with their true and empirical counterparts. Only the $k=3$ is able to capture the extremal behaviour of the underlying data.

\subsubsection{Non-exchangeable data} \label{supsubsec:asy}

\begin{figure}[h!]
    \centering
    \includegraphics[width=0.5\textwidth]{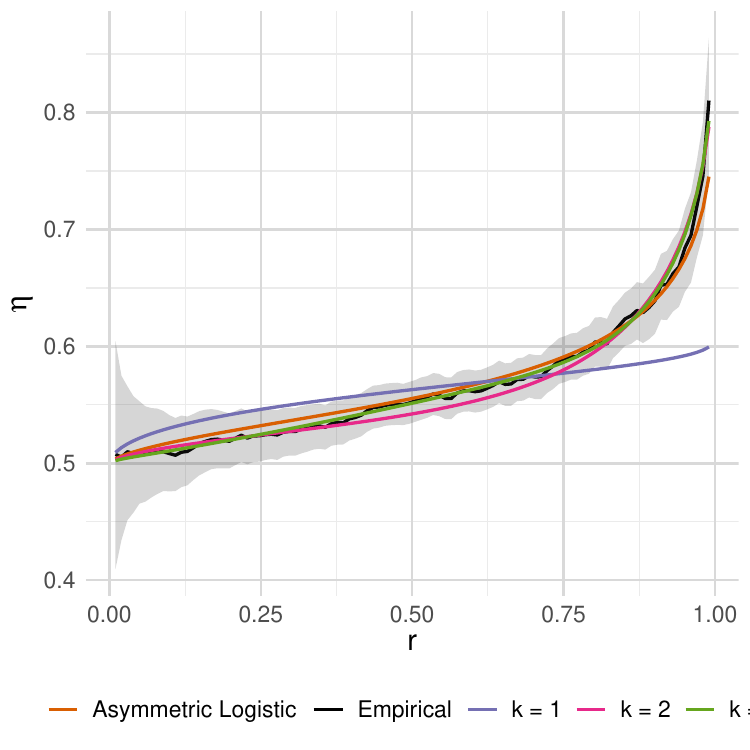}
    \caption{Estimates of $\eta_2(r)$ for $r\in (0.1)$ with true (in orange) and empirical (in black) values also shown. The pointwise $95\%$ confidence intervals for the empirical $\eta_2(r)$ are obtained through bootstrap.}
    \label{fig:asysup}
\end{figure}

Figure~\ref{fig:asysup} shows the results for $\eta_2(r)$ in the case where the underlying data exhibits asymmetry patterns given in the corresponding section of the main paper.

\subsubsection{Asymptotically independent data} \label{supsubsec:aidata}

\begin{figure}[h!]
    \centering
    \includegraphics[width = 0.9\textwidth]{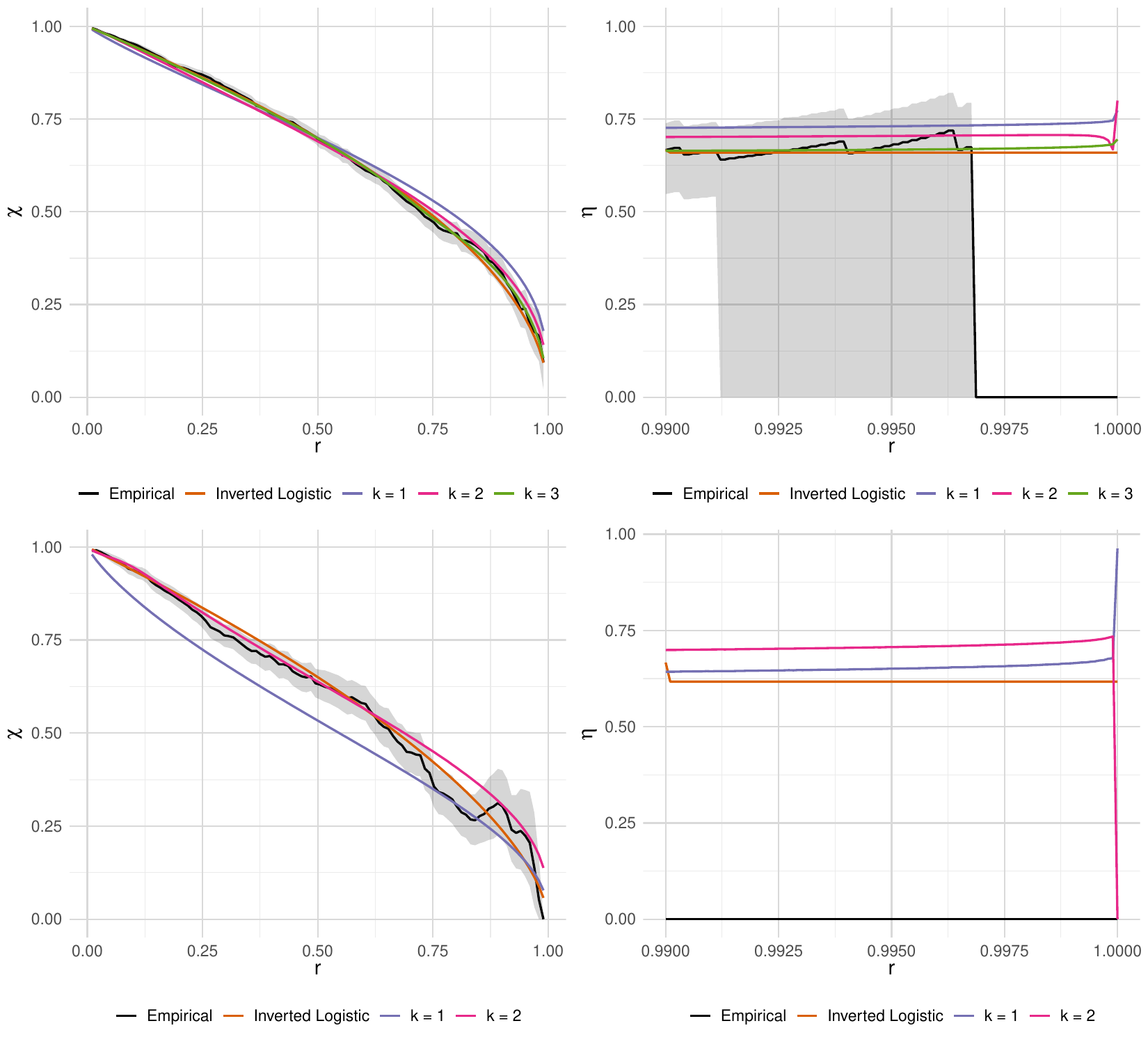}
    \caption{Estimates of $\chi_D(r)$ for $r\in (0.1)$ with true (in orange) and empirical (in black) values also shown. The corresponding results for $\eta_D(r)$ are zoomed in for $r\in[0.99,1)$ on the right. The pointwise $95\%$ confidence intervals for the empirical $\chi_D(r)$ are obtained through bootstrap. When $d=2$ (top), models with $k = 1,2$ and $3$ mixture components are considered, whereas when $d=5$ (bottom) models with only $k=1$ and $2$ mixture components are studied.}
    \label{fig:ai}
\end{figure}

\begin{figure}[h!]
    \centering
    \includegraphics[width = 0.9\textwidth]{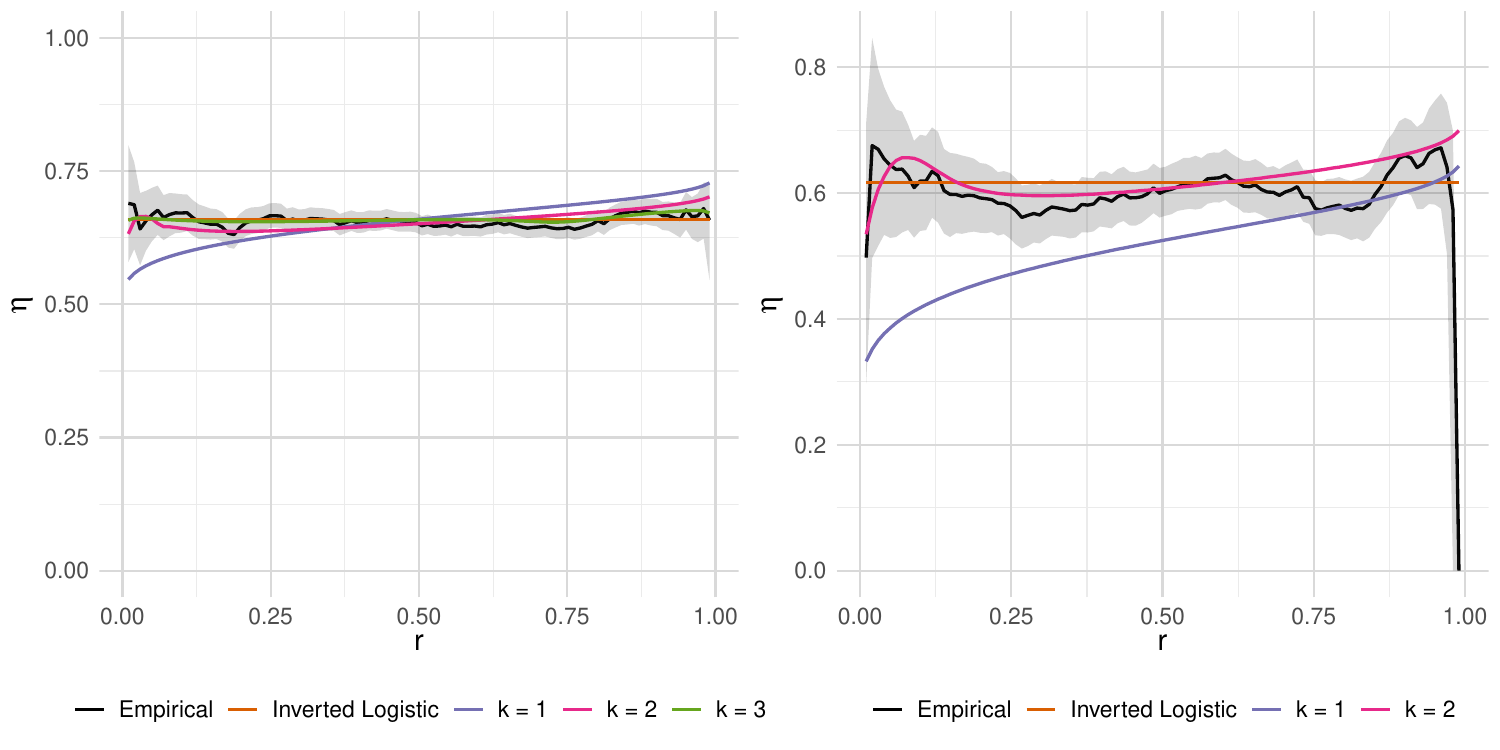}
    \caption{Estimates of $\eta_D(r)$ for $r\in (0.1)$ with true (in orange) and empirical (in black) values also shown. The pointwise $95\%$ confidence intervals for the empirical $\eta_D(r)$ are obtained through bootstrap. When $d=2$ (left), models with $k=1-3$ mixture components are considered, whereas when $d=5$ only models with $k=1-2$ mixture components are studied.}
    \label{fig:aieta}
\end{figure}

\begin{figure}[H]
    \centering
    \includegraphics[width = 0.9\textwidth]{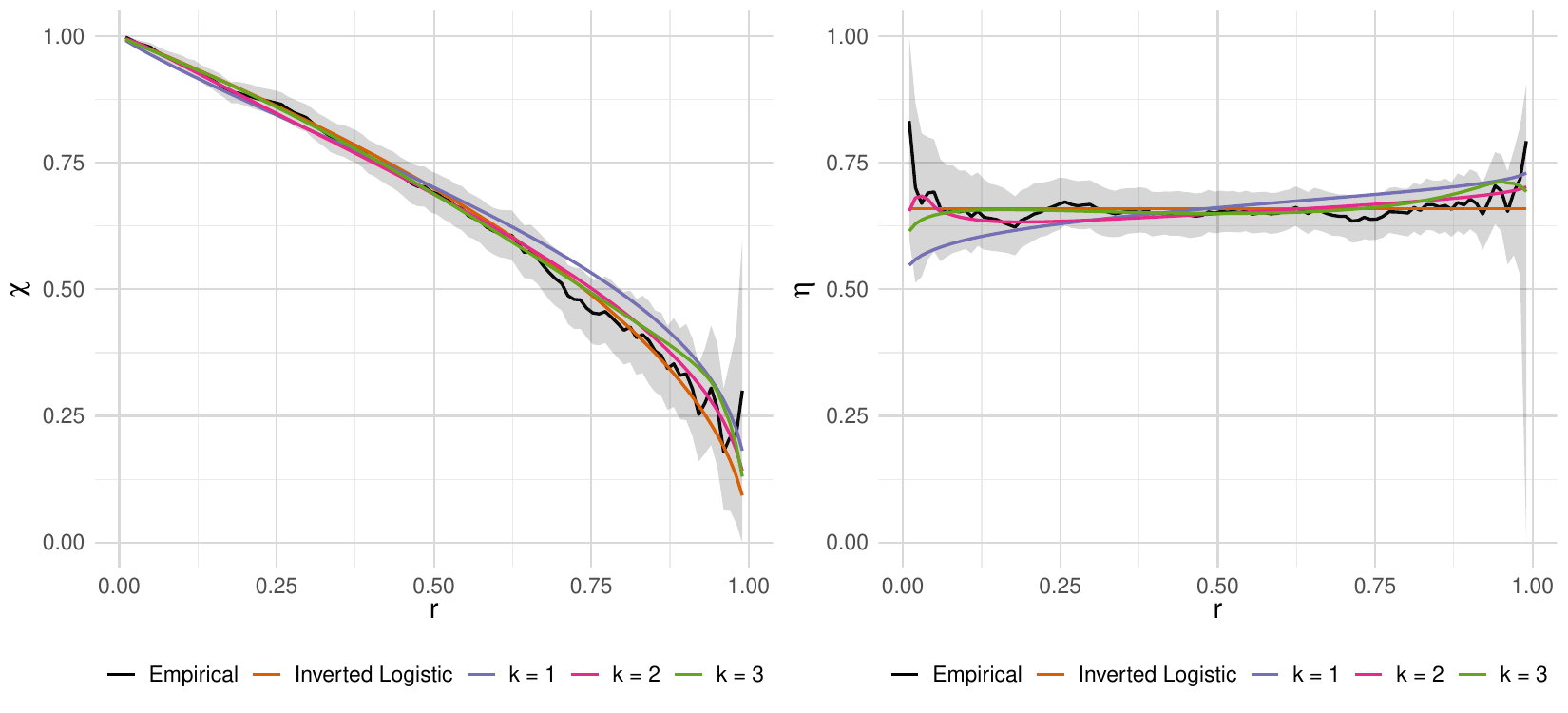}
    \caption{Estimates of $\chi_2(r)$ (left) and of $\eta_2(r)$ (right) for $r\in (0.1)$ with true (in orange) and empirical (in black) values also shown for $n=1000$. The pointwise $95\%$ confidence intervals for the empirical $\chi_2(r)$ and $\eta_2(r)$ are obtained through bootstrap.}
    \label{fig:aisup}
\end{figure}

Complementary to Section~\ref{section:modelfit} from the main paper, we assess the performance of the Gaussian mixture copula on bivariate data generated from a bivariate inverted extreme value copula with logistic dependence structure, parameter $\alpha_{IL}=0.6$ and $n=5000.$ This copula has $\chi_D=0$ and so exhibits AI. And, similarly, we consider Gaussian mixture copulas with $k=1, 2$ and $3$ mixture components. Not surprisingly given the AI nature of the underlying data, all the three specifications provide good fits even though the fitted model does not contain the true copula class as a special case. The decrease in AIC with $k>1$ in relation to when $k=1$ is $-171.77$ for $k=2$ and $-222.06$ for $k=3,$ which indicates the best fit over $k=1, 2,3$ is given by the copula with $k=3$ components. The dependence measure $\chi_D(r)$ computed from the three model fits for $r\in (0,1)$ is shown in the top left panel of Figure~\ref{fig:ai}, where a comparison with the true $\chi_2(r)$ is given. In addition, we present the results for $\eta_2(r)$ zoomed in for $r\in [0.99, 1)$ in the top right panel. There are differences, though small, between the three fits with $k=1$ slightly over-estimating the empirical and true $\chi_2(r)$ for higher values of $r$. Given that the three models seem to capture the joint behaviour for all $r,$ it can be argued that it is sufficient to consider the simplest model configuration. However, the closeness of fit for $\chi_2(r)$ may not be representative of other joint distribution characteristics, given the clear differences in AIC values. The plot for $\eta_2(r)$ across all $r,$ given in Figure~\ref{fig:aieta}, shows similar findings. We also consider a smaller sample size ($n=1000$) with $k=1,2,3$, where the results shown in Figure~\ref{fig:aisup} indicate a very good fit for all $k=1, 2$ and $3$ mixtures.

We also study the $d=5$ case with $n=1000$ and a dependence parameter of $\alpha_{IL}=0.3,$ where only $k=1$ and $2$ mixture components are consider. With a decrease in AIC of $-1221.46$ for $k=2$ in relation to $k=1,$ the model with $k=2$ is the preferred one to fit the data. This is also visible in the bottom left panel of Figure~\ref{fig:ai} with the $k=2$ model capturing the joint tail behaviour well for all levels $r\in (0,1),$ whilst with $k=1$ the model under-estimates the empirical and true $\chi_5(r)$ measures for levels $r<0.75.$ We can see from the plot for $\eta_D(r)$ for $r\in [0.99, 1)$ on the bottom right, however, that the $k=1$ model is closer to the true $\eta_5(r)$ as $r\to 1$. The results for $\eta_5(r)$ across all $r$ is given in Figure~\ref{fig:aieta}. In both studies, the $\eta_D(r)$ plots given in the right panel of Figure~\ref{fig:ai} show that for values $r$ very close to 1, the empirical estimates fail to characterise the joint behaviour, whereas the Gaussian mixture copulas with $k=1, 2, 3$ components are all able to extrapolate far into the tail. This sudden drop of the empirical estimates and their pointwise confidence intervals for $r> 0.966$ and $r>0.99$ for $d=2$ and $d=5,$ respectively, is due to the lack of observations that are jointly bigger than $r,$ resulting in $\eta_D(r)$ not being defined.

When considering a smaller sample size $(n=1000),$ the decrease in AIC with $k=3$ in relation to when $k=1$ is of $-32.15,$ with $k=2$ of $-29.47$ relative to $k=1.$ These results indicate that either the $k=2$ or the $k=3$ model is suitable to model the data, with a slight preference for the $k=3$ model. Figure~\ref{fig:aisup} shows a comparison between model-based $\chi_2(r)$ and $\eta_2(r)$ with their true and empirical counterparts. Although small, there are differences between the three fits, especially for the $k=3$ model.

\subsubsection{Weighted copula model}  \label{supsubsec:wcm}

\begin{figure}[h!]
    \centering
    \includegraphics[width=0.9\textwidth]{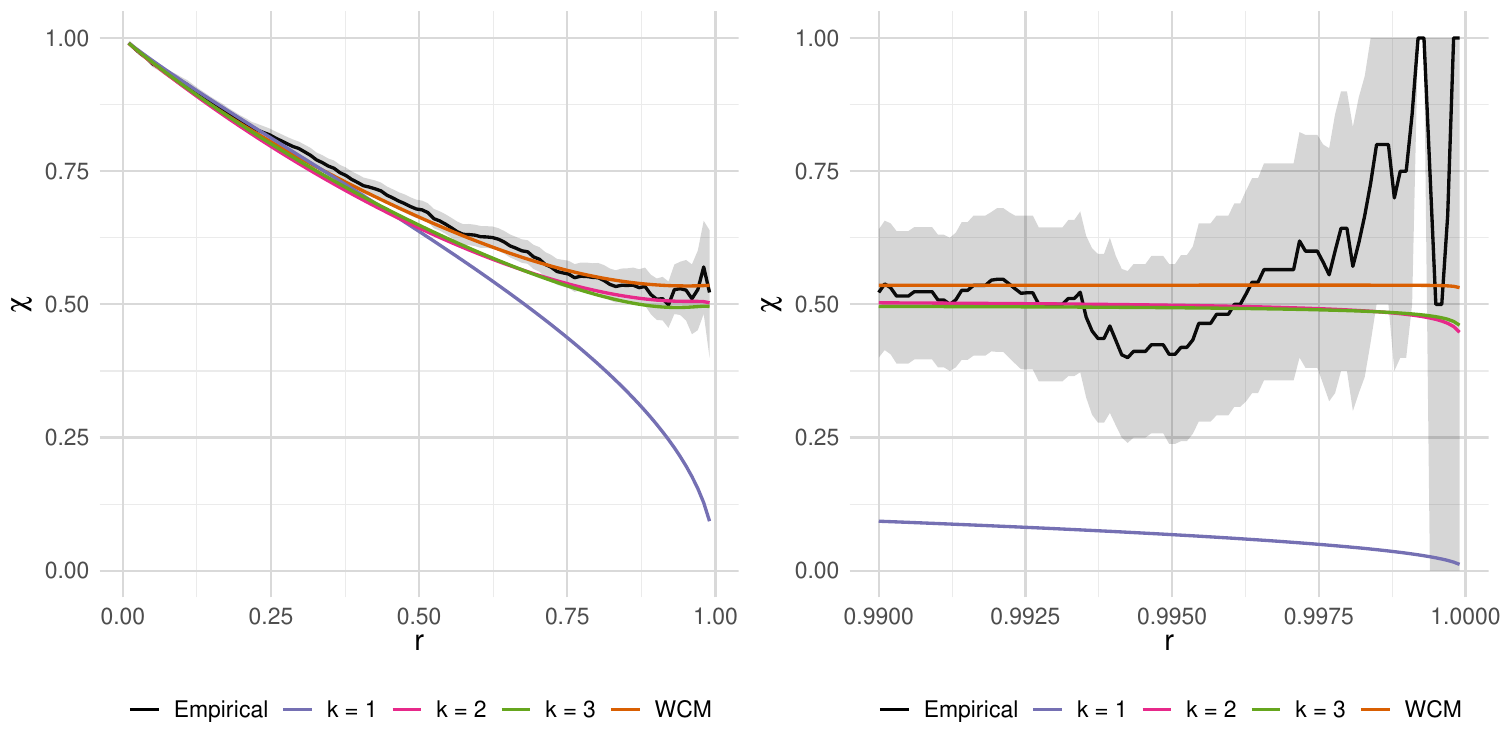}
    \caption{Estimates of $\chi_2(r)$ for $r\in (0.1)$ with true (in orange) and empirical (in black) values also shown. These are zoomed in for $r\in[0.99,1)$ on the right. The pointwise $95\%$ confidence intervals for the empirical $\chi_2(r)$ are obtained through bootstrap.}
    \label{fig:wcm}
\end{figure}

\begin{figure}[H]
    \centering
    \includegraphics[width=0.5\textwidth]{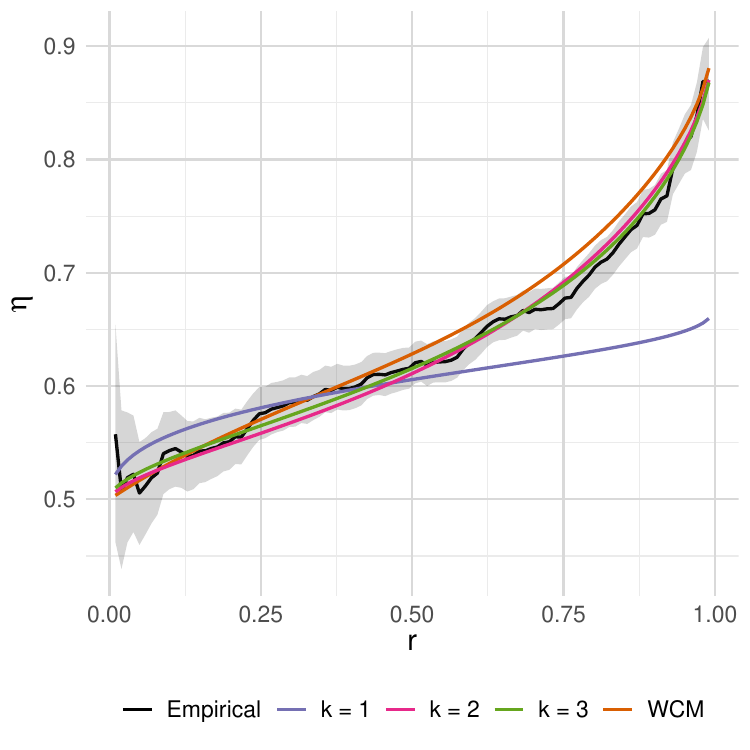}
    \caption{Estimates of $\eta_2(r)$ for $r\in (0.1)$ with true (in orange) and empirical (in black) values also shown. The pointwise $95\%$ confidence intervals for the empirical $\eta_2(r)$ are obtained through bootstrap.}
    \label{fig:wcmsup}
\end{figure}

As a final study, we assess the fit of the Gaussian mixture copula in more complex type data. To do so, we consider data generated from a configuration of the Weighted copula model (WCM). In particular, we take the copula tailored to the tail, $c_t,$ to be a bivariate extreme value copula with logistic dependence structure with dependence parameter $\alpha_L=0.3,$ and the copula tailored to the body, $c_b,$ to be a Frank copula \citep{Frank1979} with parameter $\alpha_F=2.$ Furthermore, we use the dynamic weighting function $\pi(\bm v;\theta)=(v_1v_2)^\theta,$ $\bm v=(v_1, v_2)\in [0,1]^2,$ with $\theta = 1.5,$ and $n=5000.$ Similarly to the previous cases, the decrease in AIC with $k>1$ relative to when $k=1$ is $-974.27$ for $k=2$ and $-1017.67$ for $k=3,$ indicating that the $k=3$ model provides the best fit to the data. Likewise to the AI case, the difference in AIC between the $k=2$ and $k=3$ models is very small, meaning that the Gaussian mixture copula with $k=2$ may be sufficient to fit the underlying data. This is also in agreement with the results for $\chi_2(r)$ shown in Figure~\ref{fig:wcm}, and for $\eta_2(r)$ given in Figure~\ref{fig:wcmsup}. While the $k=1$ model clearly under-estimates $\chi_2(r)$ from $r>0.5,$ the Gaussian mixture copulas with $k=2$ and $k=3$ lie closely to the true $\chi_2(r)$ for the full distribution. Moreover, the results for $\chi_2(r)$ shown in the right panel of Figure~\ref{fig:wcmsup} indicate that the empirical estimates, and their pointwise confidence intervals, become uninformative, and therefore unreliable, for $r>0.9975.$ This is not the case for the $k=2$ and $k=3$ models, for which the estimates for $\chi_2(r)$ remain stable and close to the truth.

\section{Case study: air pollution data} \label{supsec:casestudy}

\begin{figure}[h!]
    \centering
    \includegraphics[width=\textwidth]{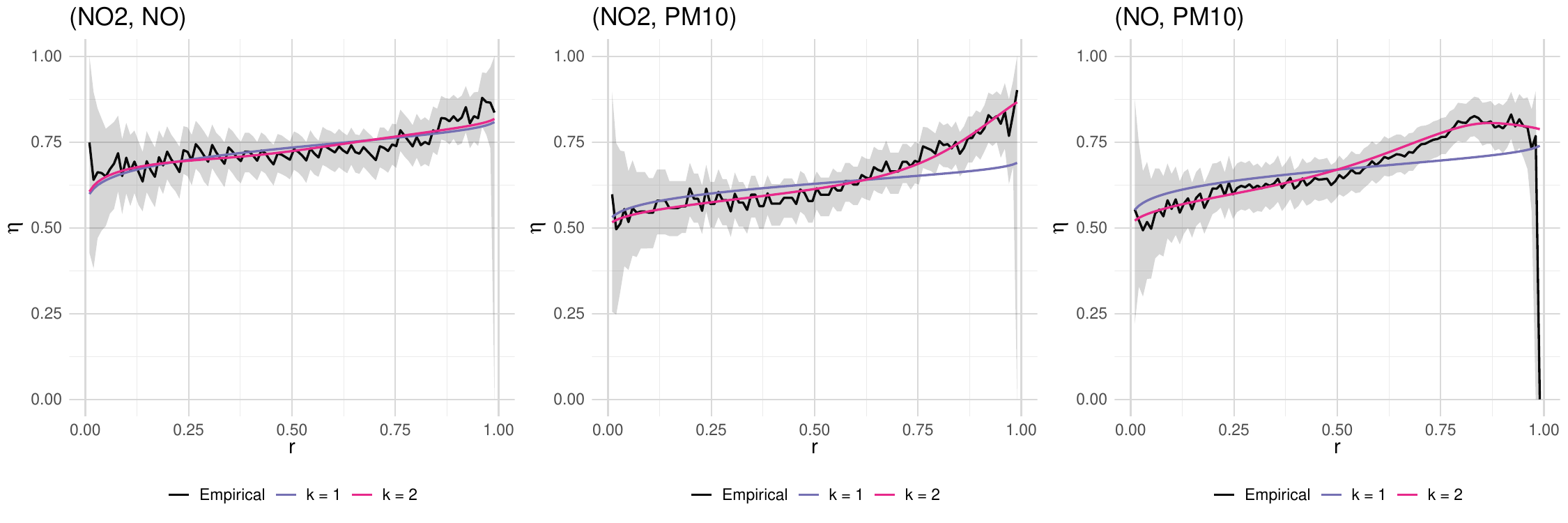}
    \caption{Estimates of $\eta_2(r)$ for $r\in (0.1)$ with empirical (in black) values also shown for pairs $(NO_2, NO)$ (left), $(NO_2, PM_{10})$ (middle) and $(NO, PM_{10})$ (right). The pointwise $95\%$ confidence intervals for the empirical $\eta_2(r)$ are obtained through bootstrap.}
    \label{fig:pairwisesup}
\end{figure}

Figure~\ref{fig:pairwisesup} shows the results for $\eta_2(r)$ for the pairwise analysis presented in the main paper.

\begin{figure}[h!]
    \centering  
    \begin{subfigure}[b]{\textwidth}
        \centering
        \includegraphics[width=0.5\textwidth]{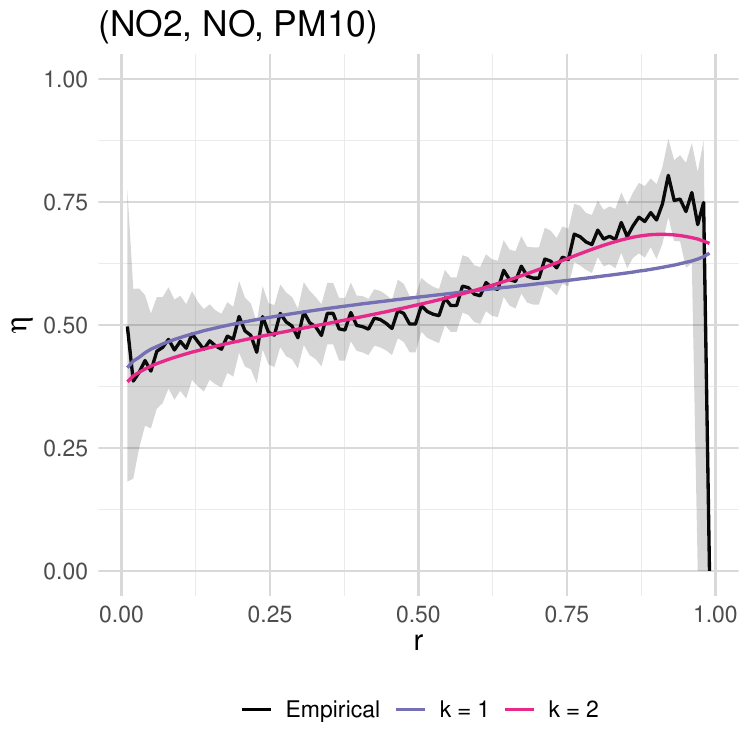}
        \caption{Estimates of $\eta_3(r)$ for $r\in (0.1)$ with empirical (in black) values also shown for the triple $(NO_2, NO, PM_{10})$. The pointwise $95\%$ confidence intervals for the empirical $\eta_3(r)$ are obtained through bootstrap.}\label{fig:appd3sup}
    \end{subfigure}
    \hfill
    \begin{subfigure}[b]{\textwidth}
        \includegraphics[width=\textwidth]{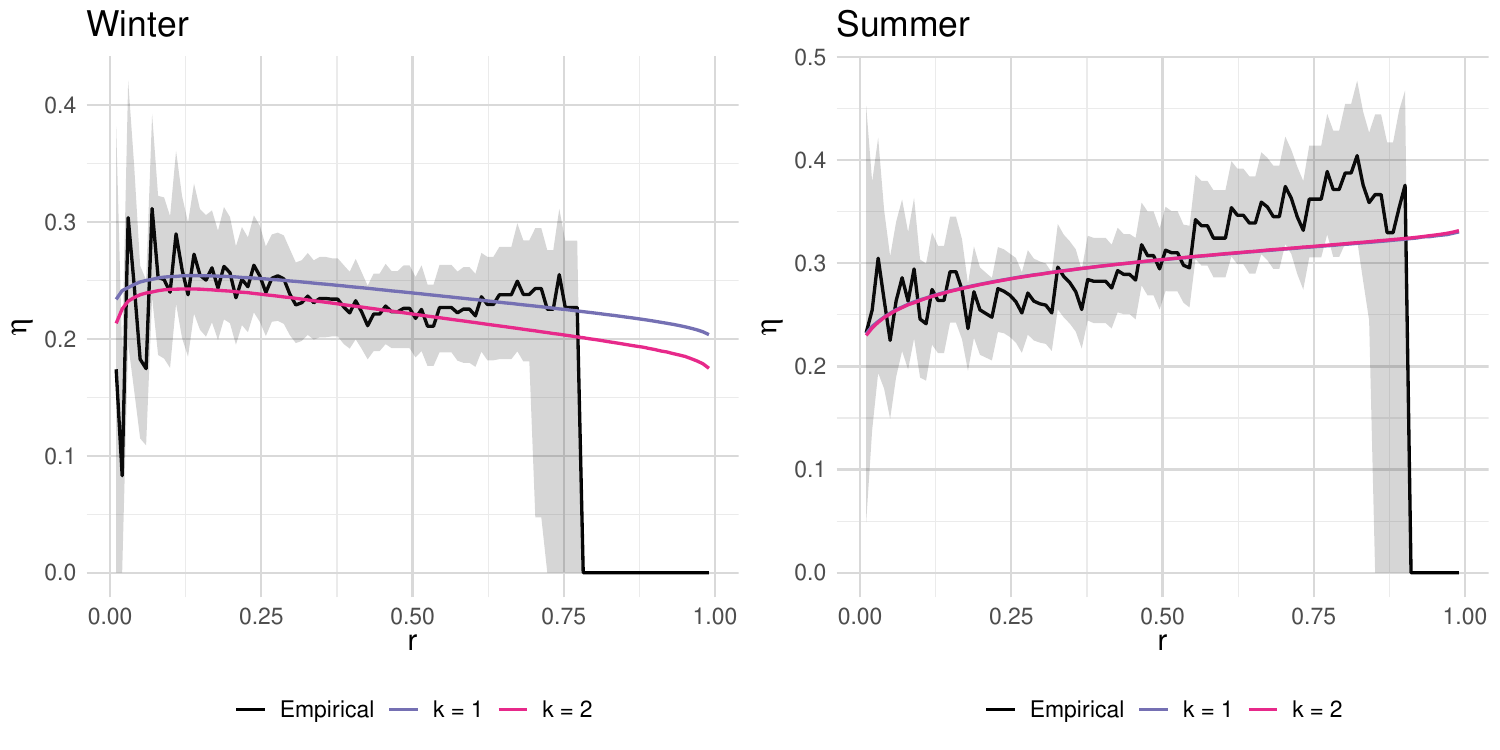}
        \caption{Estimates of $\eta_5(r)$ for $r\in (0.1)$ with empirical (in black) values also shown for $(O_3, NO_2, NO, SO_2, PM_{10})$ in the winter season (left) and the summer season (right). The pointwise $95\%$ confidence intervals for the empirical $\eta_5(r)$ are obtained through bootstrap. Note that $\eta_5(r)$ for $k=1$ and $k=2$ overlap in the right panel.}
    \label{fig:appd5sup}
    \end{subfigure}
\end{figure}

\newpage
Figure~\ref{fig:appd3sup} and \ref{fig:appd5sup} show the results, respectively, for $\eta_3(r)$ for the trivariate analysis and $\eta_5(r)$ for the full analysis presented in the corresponding section of the main paper.

\end{document}